\documentclass[12pt]{article}
\usepackage{lscape}
\usepackage{citesort}
\usepackage{amssymb}
\usepackage{appendix}
\usepackage{multirow}

\begin{document}
\def\qq{\langle \bar q q \rangle}
\def\uu{\langle \bar u u \rangle}
\def\dd{\langle \bar d d \rangle}
\def\sp{\langle \bar s s \rangle}
\def\GG{\langle g_s^2 G^2 \rangle}
\def\Tr{\mbox{Tr}}
\def\figt#1#2#3{
        \begin{figure}
        $\left. \right.$
        \vspace*{-2cm}
        \begin{center}
        \includegraphics[width=10cm]{#1}
        \end{center}
        \vspace*{-0.2cm}
        \caption{#3}
        \label{#2}
        \end{figure}
	}
	
\def\figb#1#2#3{
        \begin{figure}
        $\left. \right.$
        \vspace*{-1cm}
        \begin{center}
        \includegraphics[width=10cm]{#1}
        \end{center}
        \vspace*{-0.2cm}
        \caption{#3}
        \label{#2}
        \end{figure}
                }

\def\ds{\displaystyle}
\def\beq{\begin{equation}}
\def\eeq{\end{equation}}
\def\bea{\begin{eqnarray}}
\def\eea{\end{eqnarray}}
\def\beeq{\begin{eqnarray}}
\def\eeeq{\end{eqnarray}}
\def\ve{\vert}
\def\vel{\left|}
\def\ver{\right|}
\def\nnb{\nonumber}
\def\ga{\left(}
\def\dr{\right)}
\def\aga{\left\{}
\def\adr{\right\}}
\def\lla{\left<}
\def\rra{\right>}
\def\rar{\rightarrow}
\def\lrar{\leftrightarrow}  
\def\nnb{\nonumber}
\def\la{\langle}
\def\ra{\rangle}
\def\ba{\begin{array}}
\def\ea{\end{array}}
\def\tr{\mbox{Tr}}
\def\ssp{{\Sigma^{*+}}}
\def\sso{{\Sigma^{*0}}}
\def\ssm{{\Sigma^{*-}}}
\def\xis0{{\Xi^{*0}}}
\def\xism{{\Xi^{*-}}}
\def\qs{\la \bar s s \ra}
\def\qu{\la \bar u u \ra}
\def\qd{\la \bar d d \ra}
\def\qq{\la \bar q q \ra}
\def\gGgG{\la g^2 G^2 \ra}
\def\GG{\langle g_s^2 G^2 \rangle}
\def\g5{\gamma_5 \not\!q}
\def\x{\gamma_5 \not\!x}
\def\g5{\gamma_5}
\def\sb{S_Q^{cf}}
\def\sd{S_d^{be}}
\def\su{S_u^{ad}}
\def\sbp{{S}_Q^{'cf}}
\def\sdp{{S}_d^{'be}}
\def\sup{{S}_u^{'ad}}
\def\ssp{{S}_s^{'??}}

\def\sig{\sigma_{\mu \nu} \gamma_5 p^\mu q^\nu}
\def\fo{f_0(\frac{s_0}{M^2})}
\def\ffi{f_1(\frac{s_0}{M^2})}
\def\fii{f_2(\frac{s_0}{M^2})}
\def\O{{\cal O}}
\def\sl{{\Sigma^0 \Lambda}}
\def\es{\!\!\! &=& \!\!\!}
\def\ap{\!\!\! &\approx& \!\!\!}
\def\ar{&+& \!\!\!}
\def\arrr{\!\!\!\! &+& \!\!\!}
\def\ek{&-& \!\!\!}
\def\kek{\!\!\!\!&-& \!\!\!}
\def\cp{&\times& \!\!\!}
\def\se{\!\!\! &\simeq& \!\!\!}
\def\eqv{&\equiv& \!\!\!}
\def\kpm{&\pm& \!\!\!}
\def\kmp{&\mp& \!\!\!}
\def\mcdot{\!\cdot\!}
\def\erar{&\rightarrow&}


\def\simlt{\stackrel{<}{{}_\sim}}
\def\simgt{\stackrel{>}{{}_\sim}}


\title{
         {\Large
                 {\bf
Magnetic moments of negative-parity baryons in QCD
                 }
         }
      }

\author{\vspace{1cm}\\
{\small T. M. Aliev \thanks {
taliev@metu.edu.tr}~\footnote{Permanent address: Institute of
Physics, Baku, Azerbaijan.}\,\,,
M. Savc{\i} \thanks
{savci@metu.edu.tr}} \\
{\small Physics Department, Middle East Technical University,
06531 Ankara, Turkey }}

\date{}

\begin{titlepage}
\maketitle
\thispagestyle{empty}

\begin{abstract}

Using the most general form of the interpolating current for the octet
baryons, the magnetic moments of the negative-parity baryons are calculated
within the light-cone sum rules. The contributions coming from diagonal
transitions of the positive-parity baryons, and also from non-diagonal
transition between positive and negative-parity baryons are eliminated by
considering the combinations of different sum rules corresponding to the
different Lorentz structures. A comparison of our results on magnetic
moments of the negative-parity baryons with the other approaches existing in
literature is presented.

\end{abstract}

~~~PACS numbers: 11.55.Hx, 13.40.Gp, 14.20.Gk

\end{titlepage}

\section{Introduction}

The study of the spectroscopy, properties and structure of hadrons plays a
critical role in understanding the strong interaction at low energies. Study
of the physics of negative-parity baryons in this direction receives special
attention. There is yet very limited experimental information about 
negative-parity hyperons. Comprehensive studies on theoretical and
experimental sides can shed light in understanding dynamics of the negative-parity
baryons. For example, the mass difference of the positive and
negative-parity baryons can be attributed to the spontaneous breaking of the
chiral symmetry. In QCD sum rules approach \cite{Rmag01} the spontaneous
breakdown of the chiral symmetry is related with the condensates of
chiral-odd operators (see for example \cite{Rmag02}).

The measurement of the magnetic moment of the negative-parity baryons can
give useful information about their inner structure. Some number of
photo and electro-production experiments are planned to measure the magnetic
moments of these baryons at Mainz Microton facility
\cite{Rmag03,Rmag04}, and Jefferson
Laboratory \cite{Rmag05}. The magnetic moments of negative-parity
baryons have already been calculated in framework of the chiral and
non-relativistic constituent quark models \cite{Rmag06},
lattice QCD \cite{Rmag07}, simple quark model
\cite{Rmag08}, unitarized chiral perturbation theory \cite{Rmag09}, and effective
Hamiltonian approach \cite{Rmag10}. However, there are drastic differences
among the predictions of the above-mentioned approaches. Therefore,
motivated partly by the experimental studies, there appears the necessity
to calculate the magnetic moments of these baryons in framework of the approaches
other than listed above. In the present work we calculate the magnetic
moment of the negative-parity baryons in framework of the light-cone QCD sum
rules. The main advantage of the QCD sum rules compared to the other
approaches is that it is based on fundamental QCD Lagrangian, and it takes
into account the non-perturbative nature of the QCD vacuum. The light-cone
version of the QCD sum rules is based on the operator product expansion
(OPE) near the light-cone. This expansion is performed over the twists of
the operator rather than the dimension of the operators, as is the case in
the traditional QCD sum rules. The non-perturbative dynamics is described by
the light-cone distribution amplitudes, which appear in the matrix elements
of the nonlocal operators between the vacuum and corresponding one-particle
states. This method have successfully applied to wide range of problems in
hadron physics (for more about this method, see \cite{Rmag11}).

The structure of the present work as as follows. In section 2 we formulate
and derive the light-cone sum rules for the magnetic moments of the
negative-parity octet baryons. In the same section we also obtain the sum rules for
the masses and residues of these baryons. In section 3 we present our numerical results
for the magnetic moment of the negative-parity baryons, and discussion and
conclusion on the obtained results. 

\section{Sum rules for the magnetic moments of negative-parity baryons}

In order to obtain light-cone QCD sum rules for the magnetic moments of the
negative-parity baryons, we start by considering the following correlation
function:
\bea
\label{emag01}
\Pi = i \int d^4 x \, e^{i p x} \lla 0 \ve T\{\eta_B(x)
\bar \eta_B(0) \} \ve 0 \rra_\gamma~,
\eea
where $\eta_B$ is the interpolating current of the corresponding baryon.
According to the QCD sum rules methodology, in order to construct sum rules
for the appropriate physical quantity, the correlation function is
calculated in terms of hadrons and quark-gluon degrees of freedom, and then
with the help of quark-hadron duality ansatz these two representations are
related by using the analytical continuation.

In calculating the correlation function from QCD side, wee need the
expressions of the interpolating currents of the baryons. The general form
of the interpolating currents of octet baryons are given in \cite{Rmag12}
(see also \cite{Rmag13}.

Let us now obtain the phenomenological representation of the correlator
function. Saturating (\ref{emag01}) with the complete set of baryons having
the same quantum numbers as the interpolating current and isolating the
contributions of the ground state, we get the following expression for the
correlator function:
\bea          
\label{emag02}
\Pi \es {\lla 0 \vel \eta_B \ver B(p_2) \rra \over (p_2^2-m_B^2)}
\lla B(p_2) \gamma(q) \vel\right. B(p_1) \rra
{\lla B(p_1) \vel \bar{\eta}_B \ver 0\rra  \over (p_1^2-m_B^2)} \nnb \\
\ar {\lla 0 \vel \eta_B \ver B^\ast (p_2) \rra \over (p_2^2-m_B^{\ast 2})}
\lla B^\ast(p_2) \gamma(q) \vel\right. B^\ast(p_1) \rra 
{\lla B^\ast(p_1) \vel \bar{\eta}_B \ver 0\rra  \over (p_1^2-m_B^{\ast 2})} \nnb \\
\ar {\lla 0 \vel \eta_B \ver B(p_2) \rra \over (p_2^2-m_B^2)}
\lla B(p_2) \gamma(q) \vel\right. B^\ast(p_1) \rra
{\lla B^\ast(p_1) \vel \bar{\eta}_B \ver 0\rra  \over (p_1^2-m_B^{\ast 2})} \nnb \\
\ar {\lla 0 \vel \eta_B \ver B^\ast (p_2) \rra \over (p_2^2-m_B^{\ast 2})}
\lla B^\ast(p_2) \gamma(q) \vel\right. B(p_1) \rra
{\lla B(p_1) \vel \bar{\eta}_B \ver 0\rra  \over (p_1^2-m_B^2)} \nnb \\
\ar \cdots
\eea
where dots correspond to the contribution of the higher states and
continuum; and $B^\ast$ is the negative-parity baryon.

The matrix elements entering into Eq. (\ref{emag02}) are defined as
\bea
\label{emag03}
\lla 0 \vel \eta \ver B(p) \rra \es \lambda_B u(p)~, \nnb \\
\lla 0 \vel \eta \ver B^\ast (p)\rra \es \lambda_{B^\ast} \gamma_5 u(p)~, \nnb \\
\lla B(p_2)\gamma(q) \vel\right. B(p_1) \rra \es e \varepsilon^\mu \bar{u}(p_2)
\Bigg[f_1 \gamma_\mu - {i\sigma_{\mu\nu} q^\nu \over 2 m_B} f_2 \Bigg]u(p_1)~, \nnb \\
\lla B^\ast(p_2) \gamma(q) \vel\right. B^\ast(p_1) \rra \es e \varepsilon^\mu \bar{u}(p_2) 
\Bigg[f_1^\ast \gamma_\mu - {i\sigma_{\mu\nu} q^\nu \over 2 m_B^\ast}
f_2^\ast \Bigg]u(p_1)~, \nnb \\
\lla B^\ast(p_2) \gamma(q) \vel\right. B(p_1) \rra \es e \varepsilon^\mu \bar{u}(p_2) 
\Bigg[f_1^T \gamma_\mu - {i\sigma_{\mu\nu} q^\nu \over m_B+m_B^\ast}
f_2^T \Bigg]\gamma_5 u(p_1)~.
\eea
The structure proportional to $\gamma_\mu$ for the diagonal transformations
contains the factor $f_1+f_2$, and at $q^2=0$ it corresponds to the magnetic
moments of the corresponding baryons. Performing summation over the spins of
the baryons for the terms proportional to $\gamma_\mu$ we get
\bea
\label{emag04}
&&A^\prime (\rlap/{p}_2 + m_B) \rlap/{\varepsilon} (\rlap/{p}_1 + m_B)
+ C^\prime  (\rlap/{p}_2 - m_{B^\ast}) \rlap/{\varepsilon} (\rlap/{p}_1 -
m_{B^\ast})+ D^\prime  (\rlap/{p}_2 - m_{B^\ast}) \rlap/{\varepsilon} (\rlap/{p}_1
+ m_B) \nnb \\
\ar  D^\prime  (\rlap/{p}_2 + m_B) \rlap/{\varepsilon} (\rlap/{p}_1 -
m_{B^\ast})~,
\eea
where
\bea
\label{emag05}
A^\prime \es {\lambda_B(t^\prime) \lambda_B(t) \over (m_B^2-p_2^2)(m_B^2-p_1^2)}
(f_1+f_2)~, \nnb \\
C^\prime \es {\lambda_{B^\ast}(t^\prime) \lambda_{B^\ast}(t) \over
(m_{B^\ast}^2-p_2^2)(m_{B^\ast}^2-p_1^2)}   
(f_1^\ast+f_2^\ast)~, \nnb \\
D^\prime \es {\lambda_B(t^\prime) \lambda_{B^\ast}(t) \over
(m_{B^\ast}^2-p_2^2)(m_B^2-p_1^2)}   
\Bigg(f_1^T+{m_{B^\ast}-m_B\over m_{B^\ast}+m_B}f_2^T\Bigg)~, \nnb \\
E^\prime \es - {\lambda_{B^\ast}(t^\prime) \lambda_B(t) \over       
(m_B^2-p_2^2)(m_{B^\ast}^2-p_1^2)}       
\Bigg(f_1^T+{m_{B^\ast}-m_B\over m_{B^\ast}+m_B}f_2^T\Bigg)~,
\eea
where $t$ and $t^\prime$ are two arbitrary parameters in interpolating currents
of the baryons. 
In order to determine the magnetic moments of the negative-parity baryons,
the contributions coming from the diagonal $B \to B$ and non-diagonal $B \to
B^\ast$ and $B^\ast \to B$ transitions should be removed, i.e., only the
terms proportional to $C^\prime$ need to be determined.
In the case of diagonal transitions the quantities $(f_1+f_2)\ve_{q^2=0}$ and
$(f_1^\ast+f_2^\ast)\ve_{q^2=0}$ correspond to the magnetic moments of the positive
and negative octet baryons in natural units. As a result we have four
equations in determining the magnetic moment of the negative-parity baryons.

As has already been noted, in constructing the sum rules for the magnetic
moments the expression for the correlation function from the QCD side is
also needed. As an example we shall present the result for the $\Sigma^{\ast
+}$ case:
\bea
\label{emag06}
\Pi^{\Sigma^+} \es 4 \epsilon_{abc}\epsilon_{a^\prime b^\prime c^\prime}
\int d^4x e^{ipx} \la \gamma(q) \ve
\sum_{\ell=1}^2 \sum_{k=1}^2 \Big\{\Gamma_2^\ell S_u^{cc^\prime} (x) \Gamma_2^k
\mbox{Tr} \Big(S_s^{bb^\prime} (x) C \Gamma_1^k S_u^{aa^\prime T}(x) 
C \Gamma_1^\ell\Big) \nnb \\ 
\ar \Gamma_2^\ell \Big[ S_u^{cc^\prime}(x) (C \Gamma_1^k)^T
S_s^{bb^\prime}(x) (C \Gamma_1^\ell)^T S_u^{aa^\prime}(x)
\Gamma_2^k + S_u^{aa^\prime}(x) (C \Gamma_1^k)^T S_s^{bb^\prime T}(x) (C
\Gamma_1^\ell)^T S_u^{cc^\prime}(x) \Gamma_2^k \nnb \\
\ar S_u^{aa^\prime}(x) \Gamma_2^k \mbox{Tr}\Big(  S_s^{bb^\prime T}(x) C    
\Gamma_1^k S_u^{cc^\prime T}(x) C \Gamma_1^\ell \Big) \Big] \Big\}
\ve 0 \ra~,
\eea
where $a,b,c,a^\prime,b^\prime,c^\prime$ are the color indices; $S_q$ is the
light quark propagator. The results for the $\Sigma^0$, $\Sigma^-$, $\Xi^0$
and $\Xi^-$ baryons can easily be obtained from the above expression of the
correlation function for $\Sigma^+$ with the help of the following
replacements:
\bea
\label{emag07}
\Pi^{\Sigma^-} \es \Pi^{\Sigma^+} (u \to d)~, \nnb \\
\Pi^{\Sigma^0} \es {1\over 2} (\Pi^{\Sigma^+} + \Pi^{\Sigma^-}) \nnb \\
\Pi^{\Xi^0}    \es \Pi^{\Sigma^+} (u \to s)~, \nnb \\
\Pi^{\Xi^-}    \es \Pi^{\Xi^0} (u \to d)~.
\eea
The expression for the light-quark operator we use in further analysis
has the following form (see for example \cite{Rmag13}):
\bea
\label{emag08}
\lefteqn{
S_q(x) = \lla 0 \vel T \left\{ \bar q(x) q(0) \right\} \ver 0 \rra}\nnb \\
\es \frac{i \not\!x}{2 \pi^2 x^4} - \frac{m_q}{4 \pi^2 x^2} - \frac{\qq}{12}
\ga 1 - \frac{i m_q}{4} \not\!x \dr - \frac{x^2}{192} m_0^2 \qq 
\ga 1 - \frac{i m_q}{6} \not\!x \dr  \nnb \\
\ek i g_s \int_0^1 dv \Bigg[ \frac{\not\!x}{16 \pi^2 x^2} G_{\mu\nu}(vx)
\sigma^{\mu\nu} - v x^\mu G_{\mu\nu} (vx) \gamma^\nu \frac{i}{4 \pi^2 x^2} \nnb \\
\ek \frac{i m_q}{32 \pi^2}  G_{\mu\nu} \sigma^{\mu\nu}
\ga \ln \frac{-x^2 \Lambda^2}{4} + 2 \gamma_E \dr \Bigg]~,
\eea
where $\Lambda$ is the energy cut off separating perturbative and
non-perturbative sectors. We keep only linear quark
mass terms in (\ref{emag08}) since the contributions coming from
higher order mass terms are negligibly small. It can also easily be seen
from the expression of the light-quark propagator that, one-gluon
operator is retained while the two-gluon $\bar{q} GG q$ and four-quark
$\bar{q}q\bar{q}q$ operators are neglected. Neglecting these operators is
justified with the help of the conformal spin expansions
\cite{Rmag14,Rmag15}.

Few words about the calculation technique are in order. If a photon interact
with the quark fields perturbatively, its contribution can be calculated by
replacing one of the propagators in Eq. (\ref{emag06}) with the following
one:
\bea
\label{emag09}       
 S = - {1\over 2} \int dy \, y_\nu {\cal F}^{\mu\nu} S^{free} (x-y) \gamma_\mu
S^{free}(y)~,
\eea
where ${\cal F}^{\mu\nu}$ is the electromagnetic field strength
tensor, and the Fock-Schwinger gauge $x_\mu A^\mu=0$ has already been taken into
account in  the relation $A_\mu = {1\over 2} {\cal F}_{\mu\nu} y^\nu$.
Note also that, $S^{free}$ in Eq. (\ref{emag09}) is the free quark operator which
corresponds to the first term on the right hand side of Eq. (\ref{emag08}),
and the following two propagators are the full propagators of
the quarks. The contribution coming from the photon's interacting
with a quark field non-perturbatively is described by replacing one of the
propagators with
\bea         
\label{nolabel}
S_{\rho\sigma}^{ab} = - {1\over 4} \bar{q}^a \Gamma_j q^b
(\Gamma_j)_{\rho\sigma}~, \nnb
\eea
where $\Gamma_j$ describes the full set of Dirac matrices; and the two
remaining propagators are described by Eq. (\ref{emag08}).

Substituting Eq. (\ref{emag09}) into (\ref{emag06}) we see that
the contributions corresponding to the non-perturbative interaction
of a photon with the quark field appear in the matrix elements of the
nonlocal operators $\bar{q} \Gamma_j q$, $\bar{q} G_{\mu\nu} \Gamma_j q$ between the
vacuum and photon states, i.e., $\lla \gamma \vel \bar{q} \Gamma_j q \ver 0 \rra$
and $\lla \gamma \vel \bar{q}G_{\mu\nu} \Gamma_j q \ver 0 \rra$, respectively.
These matrix elements are determined in terms of the photon distribution
amplitudes (DAs)which are given in \cite{Rmag16}.

Using Eqs. (\ref{emag06}) and (\ref{emag08}), and the definitions of DAs of
a photon, one can calculate the theoretical part of the correlation
function. We present the explicit expressions of invariant functions
$\Pi_1$, $\Pi_2$, $\Pi_3$ and $\Pi_4$ corresponding to the structures
$\rlap/{p}\rlap/{\varepsilon}\rlap/{q}$,
 $\rlap/{p}\rlap/{\varepsilon}$,
$\rlap/{\varepsilon}\rlap/{q}$ and $\rlap/{\varepsilon}$ for the
$\Sigma^{0\ast}$ baryon, respectively.
The results for the $\Lambda^\ast$ baryon can be obtained by using the following
relation between the interpolating currents of $\Sigma^{0\ast}$ and
$\Lambda^\ast$ baryons \cite{Rmag17}:
\bea
\label{nolabel}
2 \eta_{\Sigma^0} (d \lrar s) = - \eta_{\Sigma^0} - \sqrt{3}
\eta_{\Lambda}~. \nnb
\eea

The final step in deriving the sum rules for the magnetic moment of the
negative-parity baryons is performing double Borel transformation on the
variables $p_1^2=(p+q)^2$ and $p_2^2=p^2$ on both sides of the correlation
function and equating the coefficients of the corresponding structures,
the result of which is
\bea
\label{emag10}
A + C + D + E \es \Pi_1^{Bor}~, \nnb \\
m_B A - m_{B^\ast} C - m_{B^\ast} D + m_B E \es \Pi_2^{Bor}~, \nnb \\
(m_B A + m_{B^\ast}) (D - E) \es \Pi_3^{Bor}~, \nnb \\
m_B^2 A + m_{B^\ast}^2 C - m_B m_{B^\ast} (D + E) \es \Pi_4^{Bor}~, \nnb \\
\eea
where
\bea
\label{nolabel}
A \es \lambda_B(t^\prime) \lambda_B(t) e^{-{m_B^2 \over M_1^2} - {m_B^2\over
M_2^2}} (f_1+f_2)~,\nnb \\
C \es \lambda_{B^\ast}(t^\prime) \lambda_{B^\ast}(t) e^{-{m_{B^\ast}^2 \over M_1^2}
- {m_{B^\ast}^2\over M_2^2}} (f_1^\ast+f_2^\ast)~,\nnb \\
D \es \lambda_{B^\ast}(t^\prime) \lambda_B(t) e^{-{m_B^2 \over M_1^2} -
{m_{B^\ast}^2 \over M_2^2}} \Bigg(f_1^T+{m_{B^\ast}-m_B\over
m_{B^\ast}+m_B}f_2^T\Bigg)~, \nnb \\
E \es - \lambda_B(t^\prime) \lambda_{B^\ast}(t) e^{-{m_{B^\ast}^2 \over M_1^2} -
{m_B^2 \over M_2^2}} \Bigg(f_1^T+{m_{B^\ast}-m_B\over               
m_{B^\ast}+m_B}f_2^T\Bigg)~.\nnb
\eea

The sum rules for the magnetic moments of the negative-parity baryons is
obtained by solving Eq. (\ref{emag10}) for $C$ in terms of the invariant
functions $\Pi_i$. Since $C$ corresponds to the diagonal transition, we set
$t^\prime=t$ in the expressions of the interpolating currents. The resulting
expression of the magnetic moment of the negative-parity baryons is
\bea
\label{emag11}   
\mu_{B^\ast} \es {e^{m_{B^\ast}^2/M^2} \over \lambda_{B^\ast}^2
(m_B+m_{B^\ast}) (m_B^2+3 m_{B^\ast}^2)} \Big\{
\Big[ m_B (m_B-m_{B^\ast}) - 2 m_{B^\ast}^2 \Big] \Pi_1^{Bor}\nnb \\
\ek 2 m_B (m_B+m_{B^\ast}) \Pi_2^{Bor} -
(m_B-3 m_{B^\ast}) \Pi_3^{Bor} - m_B (m_B+m_{B^\ast}) \Pi_4^{Bor} \Big\}~,
\eea 
where we have used $M_1^2=2 M^2$, $M_2^2=2 M^2$. The expressions of
$\Pi_i^{Bor}$ are rather lengthy, and therefore we do not present them in
the present work.

It follows from Eq. (\ref{emag11}) that in determining numerical values of 
the magnetic moments of the negative-parity baryons the corresponding
residues are needed. These residues can be obtained from the following
correlator:
\bea
\label{emag12}
\Pi(p^2) = i \int d^4x e^{ipx} \lla 0 \vel T\Big\{ \eta_B(x) \bar{\eta}_B (0)
\ver 0 \rra~,
\eea
which can be written in terms of the invariant functions as
\bea
\label{nolabel}
\Pi(p^2) = \Pi_1\rlap/{p} + \Pi_2 I~. \nnb
\eea
Saturating (\ref{emag12}) with the corresponding positive and negative
baryons, we get
\bea
\label{emag13}
\Pi(p^2) = {\vel\lambda_{B^\ast} \ver^2 \over m_{B^\ast}^2-p^2}
(\rlap/{p}-m_{B^\ast}) + {\vel\lambda_B \ver^2 \over
m_B^2-p^2} (\rlap/{p}+m_B)~.
\eea
By eliminating the contributions of the positive-parity baryon
from these equations and performing Borel transformation over the variable
$p^2$, for the mass and residues, we get
\bea
\label{nolabel}
m_{B^\ast}^2 \es {\ds \int_{s_0}^s s \, ds e^{-s/M^2} \Big[m_B \mbox{Im} \Pi_1(s)
- \mbox{Im} \Pi_2(s) \Big] \over 
\ds \int_{s_0}^s ds e^{-s/M^2} \Big[m_B \mbox{Im} \Pi_1(s)  
- \mbox{Im} \Pi_2(s) \Big]} \nnb \\
\vel \lambda_{B^\ast} \ver^2 \es {e^{m_{B^\ast}^2/M^2} \over m_B+m_{B^\ast} }
{1\over \pi} \int ds e^{-s/M^2} \Big[ m_B \mbox{Im} \Pi_1(s)    
- \mbox{Im} \Pi_2(s) \Big] \nnb~.
\eea
The spectral densities $\mbox{Im} \Pi_1(s)$ and $\mbox{Im} \Pi_2(s)$ 
are calculated in \cite{Rmag13}, and for this reason we do not present them
in this work.

\section{Numerical analysis}

In this section we perform numerical calculations of the sum rules for
magnetic moments of the negative-parity baryons. The main non-perturbative
input parameters of the light-cone QCD sum rules for the magnetic moments
are the photon DAs which are all given in \cite{Rmag16}.

The values of the other input parameters are
$\qu~(1~GeV)=\qd~(1~GeV)=-(0.243)^3~GeV^3$, $\qs~(1~GeV)=0.8\qu~(1~GeV)$,
$m_0^2 = (0.8\pm0.2)~GeV^2$ \cite{Rmag18}, $\Lambda=(0.5\pm1.0)~GeV$
\cite{Rmag19}, and $f_{3\gamma}=-0.039$ \cite{Rmag16}, the magnetic
susceptibility $\chi~(1~GeV)=-2.85 \pm0.5~GeV^{-2}$ \cite{Rmag20} and
$m_s~(2~GeV)=111 \pm6~MeV$ \cite{Rmag21}.

The sum rules for the magnetic moments of the negative-parity baryons
contain three auxiliary parameters, namely, the arbitrary parameter $t$
appearing in interpolating current, Borel mass parameter $M^2$ and the
continuum threshold $s_0$. Obviously the magnetic
moments are expected to be independent of these parameters.  
In implementing the numerical analysis program, we first look for the region
where magnetic moments of the negative-parity baryons are independent of
$M^2$ at properly chosen values of the parameters $s_0$ and $t$. It should
be noted here that the QCD sum rules analysis restricts the arbitrary
parameter $t$ to posses only positive values. The upper bound of $M^2$ is
obtained by demanding that the continuum contribution should be less than,
say, half of the perturbative contributions. The lower bound of $M^2$ is
determined by requiring that the contributions of the highest power of
$1/M^2$ terms contribute less than $30\%$ of the highest $M^2$ terms. These
two conditions lead to the following domains:
\bea
\label{nolabel}
1.5 \le M^2 \le 2.5~GeV^2~\mbox{for $p^\ast$ and $n^\ast$, and}~\nnb\\
1.6 \le M^2 \le 3.0~GeV^2~\mbox{for $\Lambda^\ast$, $\Sigma^\ast$ and
$\Xi^\ast$}~.\nnb
\eea
For this reason as we draw figures depicting the $M^2$ dependence of the magnetic
moments of the negative-parity baryons positive values of the
arbitrary parameter $t$ have been used, and different ``working regions"
of $M^2$ are used for different members of the negative-parity octet baryons.
In these preferred regions of $M^2$ magnetic moments of the negative-parity
baryons are practically independent of $M^2$ at fixed values of $s_0$ and
$t$.
 
The next problem which needs to be solved is finding the working region
for the arbitrary parameter $t$ where again magnetic moments of the negative-parity
baryons are independent of it. For this goal the mass sum rules have been
used. We see that $\lambda_{B^\ast}^2$ is almost independent of $t$ if it
varies in the region $0.2 \le t \le 1.5$, which is common for all negative
parity baryons at fixed values of $M^2$. We shall use use this boundary for
$t$ in further numerical analysis of the magnetic moment of the negative
parity baryons. It should be noted here that when parameter $t$ varies in
this domain the mass sum rules predict the following values for the mass of
the negative-parity baryons: $m_\Lambda=1.75~GeV$, $m_\Xi=1.80~GeV$,
$m_\Sigma=1.7~GeV$, at $s_0=4.0~GeV^2$ and when the Borel parameter varies in
the region $1.4 \le M^2 \le 2.2~GeV^2$. These predictions for mass of the
the negative
parity baryons are very close to their experimental values \cite{Rmag22}.
It should be stressed here that the Borel mass parameters appearing in the
sum rules for magnetic moments and in the mass sum rules are different in
general.

The final step of our procedure is determination of the region for the
arbitrary parameter $t$ by using the domain which follows from the mass sum
rules analysis in which the sum rules exhibit good stability to the
variation of $t$ at fixed values of $M^2$. Our numerical calculations show
that the regions of $t$, where the magnetic moments of the negative-parity
baryons are independent of it, gets narrower. For example, the best stability
of the magnetic moments are achieved when $t$ lies in the region $0.9 \le t \le 1.0$
for $n^\ast$ and $p^\ast$, and $0.6 \le t \le 0.7$ for the $\Lambda^\ast$,
$\Sigma^\ast$ and $\Xi^\ast$ baryons, respectively.

The results of our numerical analysis for the negative
parity baryons are all summarized in Table I. For completeness, in
Table I, we also present the predictions of the chiral and
non-relativistic constituent quark
models \cite{Rmag06}, lattice calculations \cite{Rmag07}, simple quark model
\cite{Rmag08} and unitarized chiral perturbation theory \cite{Rmag09}, and
effective Hamiltonian approach \cite{Rmag10}. It should be noted here that
the magnetic moments of the $\Lambda(1670)$ baryon is calculated within the
framework of the unitarized chiral perturbation theory in \cite{Rmag23}
which predicts $\mu_{\Lambda^\ast} = -0.29\mu_N$, where $\mu_N$ is the
nuclear magneton. 

From this table we see that the results obtained by the above-mentioned
approaches are quite different, not only in magnitude, but also in sign in
many cases. On the other hand, our results predicts that
$\mu_{\Xi^{-\ast}}\simeq \mu_{\Sigma^{-\ast}}$, $\mu_{\Xi^{0\ast}}\simeq
\mu_{n^\ast}$, $\mu_{\Sigma^{+\ast}}\simeq \mu_{p\ast}$,
$\mu_{\Sigma^{-\ast}} + \mu_{n\ast} \simeq - \mu_{p\ast}$,
$2 \mu_{\Lambda^\ast} \simeq \mu_{n^\ast}$, which are all very 
close to the exact $SU(3)$ symmetry relation.

Naively, one can expect that the magnetic moments of the negative-parity
baryons can be obtained from the positive ones by the relation
\bea
\label{nolabel}
\vel \mu_- \ver = {m_B \over m_{B^\ast}} \vel \mu_+ \ver~. \nnb
\eea
When used, this relation leads to the following results (for magnetic
moments of the positive-parity baryons we have used their experimental
results): $\vel \mu_{p^\ast} \ver \simeq 1.7 \mu_N$, $\vel \mu_{n^\ast} \ver \simeq 1.1 \mu_N$,
$\mu_{\Sigma^{+\ast}} \simeq 1.7 \mu_N$, $\vel \mu_{\Sigma^{-\ast}} \ver \simeq 0.8 \mu_N$,
$\vel \mu_{\Sigma^{0\ast}} \ver \simeq 0.46 \mu_N$, $\vel
\mu_{\Xi^{0\ast}}\ver \simeq \mu_N$,
$\vel \mu_{\Xi^{-\ast}}\ver \simeq 0.5 \mu_N$, and $\mu_{\Lambda^\ast} \simeq -0.4
\mu_N$. When compared with these naive estimations we see
that, except for the $n^\ast$ case, our results are very close with the
above-presented naive estimations.

In summary, using the light-cone QCD sum rules method we calculate the
magnetic moments of the negative-parity octet baryons. In our analysis the
contaminations originating from the diagonal transitions of the positive
parity baryons, as well as non-diagonal transitions among negative and
positive-parity baryons are all eliminated by taking the contributions
coming from the sum rules corresponding to the different Lorentz structures
into account. Furthermore, we present a comparison of our results on the
magnetic moments of the negative-parity octet baryons with those lattice,
constituent quark model, unitarized chiral perturbation chiral perturbation
theory, and chiral constituent quark model predictions. It is observed that
the predictions of these approaches totally differ from each other in many
cases. Therefore, any future experimental measurement would be very
important for choosing the ``right" theory and understanding the structure
of these baryons. 


\begin{table}[h]

\renewcommand{\arraystretch}{1.3}
\addtolength{\arraycolsep}{-0.5pt}
\small
$$
\begin{array}{|l|c|c|c|c|c|c|c|}
\hline \hline
                      & \mbox{Our result} & \cite{Rmag06} & \cite{Rmag06}
                      & \cite{Rmag07} & \cite{Rmag08}     & \cite{Rmag09}
                      & \cite{Rmag10}                                                              \\  \hline
\mu_{p^\ast}          & 1.4    & 2.085     &   1.894   & -1.8 & 1.9       & 1.1        & 1.24      \\ 
\mu_{n^\ast}          & -0.54  & -1.569    &  -1.284   & -1   & 1.2       & -0.25      & -0.84     \\
\mu_{\Sigma^{+\ast}}  & 1.8    & 1.8       &   1.814   & -0.6 & \cdots & \cdots  & \cdots \\
\mu_{\Sigma^{0\ast}}  & 0.4    & 0.79      &   0.82    &  0.1 & \cdots & \cdots  & \cdots \\
\mu_{\Sigma^{-\ast}}  & -1.1   & -1        &  -0.689   &    1 & \cdots & \cdots  & \cdots \\
\mu_{\Xi^{0\ast}}     & -0.55  & -1.442    &  -0.99    & -0.5 & \cdots & \cdots  & \cdots \\
\mu_{\Xi^{-\ast}}     & -1.2   & -0.165    &  -0.315   & -0.8 & \cdots & \cdots  & \cdots \\
\mu_{\Lambda^\ast}    & -0.26  & \cdots & \cdots & -0.1 & -1.9      & -0.29      & \cdots \\ 
\hline \hline
\end{array}
$$
\caption{The magnetic moments of the negative-parity
octet baryons in units of nuclear magneton $\mu_N$ predicted by the light
cone QCD sum rules (our result), chiral and non-relativistic constituent
quark models \cite{Rmag06}, lattice
calculations \cite{Rmag07}, simple quark model \cite{Rmag08}, unitarized
chiral perturbation theory \cite{Rmag09}, non-relativistic constituent quark
model \cite{Rmag06}, and effective Hamiltonian \cite{Rmag10} approaches.}
\renewcommand{\arraystretch}{1}
\addtolength{\arraycolsep}{-1.0pt}
\end{table}

\section*{Acknowledgments}

We thank V. S. Zamiralov for his collaboration at the early stage of this
work and K. Azizi for useful discussions.

\newpage

\section*{Appendix A}
\setcounter{equation}{0}
\setcounter{section}{0}

For completeness, in this Appendix we present the matrix elements  $ \langle\gamma(q)\mid\bar q
 \Gamma_{i}q\mid0\rangle$ which are calculated in terms of
 the photon distribution amplitudes (DAs) \cite{Rmag10}.

 \bea
&&\langle \gamma(q) \vert  \bar q(x) \sigma_{\mu \nu} q(0) \vert  0
\rangle  = -i e_q \bar q q (\varepsilon_\mu q_\nu - \varepsilon_\nu
q_\mu) \int_0^1 du e^{i \bar u qx} \left(\chi \varphi_\gamma(u) +
\frac{x^2}{16} \mathbb{A}  (u) \right) \nnb \\ &&
-\frac{i}{2(qx)}  e_q \qq \left[x_\nu \left(\varepsilon_\mu - q_\mu
\frac{\varepsilon x}{qx}\right) - x_\mu \left(\varepsilon_\nu -
q_\nu \frac{\varepsilon x}{q x}\right) \right] \int_0^1 du e^{i \bar
u q x} h_\gamma(u)
\nnb \\
&&\langle \gamma(q) \vert  \bar q(x) \gamma_\mu q(0) \vert 0 \rangle
= e_q f_{3 \gamma} \left(\varepsilon_\mu - q_\mu \frac{\varepsilon
x}{q x} \right) \int_0^1 du e^{i \bar u q x} \psi^v(u)
\nnb \\
&&\langle \gamma(q) \vert \bar q(x) \gamma_\mu \gamma_5 q(0) \vert 0
\rangle  = - \frac{1}{4} e_q f_{3 \gamma} \epsilon_{\mu \nu \alpha
\beta } \varepsilon^\nu q^\alpha x^\beta \int_0^1 du e^{i \bar u q
x} \psi^a(u)
\nnb \\
&&\langle \gamma(q) | \bar q(x) g_s G_{\mu \nu} (v x) q(0) \vert 0
\rangle = -i e_q \qq \left(\varepsilon_\mu q_\nu - \varepsilon_\nu
q_\mu \right) \int {\cal D}\alpha_i e^{i (\alpha_{\bar q} + v
\alpha_g) q x} {\cal S}(\alpha_i)
\nnb \\
&&\langle \gamma(q) | \bar q(x) g_s \tilde G_{\mu \nu} i \gamma_5 (v
x) q(0) \vert 0 \rangle = -i e_q \qq \left(\varepsilon_\mu q_\nu -
\varepsilon_\nu q_\mu \right) \int {\cal D}\alpha_i e^{i
(\alpha_{\bar q} + v \alpha_g) q x} \tilde {\cal S}(\alpha_i)
\nnb \\
&&\langle \gamma(q) \vert \bar q(x) g_s \tilde G_{\mu \nu}(v x)
\gamma_\alpha \gamma_5 q(0) \vert 0 \rangle = e_q f_{3 \gamma}
q_\alpha (\varepsilon_\mu q_\nu - \varepsilon_\nu q_\mu) \int {\cal
D}\alpha_i e^{i (\alpha_{\bar q} + v \alpha_g) q x} {\cal
A}(\alpha_i)
\nnb \\
&&\langle \gamma(q) \vert \bar q(x) g_s G_{\mu \nu}(v x) i
\gamma_\alpha q(0) \vert 0 \rangle = e_q f_{3 \gamma} q_\alpha
(\varepsilon_\mu q_\nu - \varepsilon_\nu q_\mu) \int {\cal
D}\alpha_i e^{i (\alpha_{\bar q} + v \alpha_g) q x} {\cal
V}(\alpha_i) \nnb \\ && \langle \gamma(q) \vert \bar q(x)
\sigma_{\alpha \beta} g_s G_{\mu \nu}(v x) q(0) \vert 0 \rangle  =
e_q \qq \left\{
        \left[\left(\varepsilon_\mu - q_\mu \frac{\varepsilon x}{q x}\right)\left(g_{\alpha \nu} -
        \frac{1}{qx} (q_\alpha x_\nu + q_\nu x_\alpha)\right) \right. \right. q_\beta
\nnb \\ && -
         \left(\varepsilon_\mu - q_\mu \frac{\varepsilon x}{q x}\right)\left(g_{\beta \nu} -
        \frac{1}{qx} (q_\beta x_\nu + q_\nu x_\beta)\right) q_\alpha
\nnb \\ && -
         \left(\varepsilon_\nu - q_\nu \frac{\varepsilon x}{q x}\right)\left(g_{\alpha \mu} -
        \frac{1}{qx} (q_\alpha x_\mu + q_\mu x_\alpha)\right) q_\beta
\nnb \\ &&+
         \left. \left(\varepsilon_\nu - q_\nu \frac{\varepsilon x}{q.x}\right)\left( g_{\beta \mu} -
        \frac{1}{qx} (q_\beta x_\mu + q_\mu x_\beta)\right) q_\alpha \right]
   \int {\cal D}\alpha_i e^{i (\alpha_{\bar q} + v \alpha_g) qx} {\cal T}_1(\alpha_i)
\nnb \\ &&+
        \left[\left(\varepsilon_\alpha - q_\alpha \frac{\varepsilon x}{qx}\right)
        \left(g_{\mu \beta} - \frac{1}{qx}(q_\mu x_\beta + q_\beta x_\mu)\right) \right. q_\nu
\nnb \\ &&-
         \left(\varepsilon_\alpha - q_\alpha \frac{\varepsilon x}{qx}\right)
        \left(g_{\nu \beta} - \frac{1}{qx}(q_\nu x_\beta + q_\beta x_\nu)\right)  q_\mu
\nnb \\ && -
         \left(\varepsilon_\beta - q_\beta \frac{\varepsilon x}{qx}\right)
        \left(g_{\mu \alpha} - \frac{1}{qx}(q_\mu x_\alpha + q_\alpha x_\mu)\right) q_\nu
\nnb \\ &&+
         \left. \left(\varepsilon_\beta - q_\beta \frac{\varepsilon x}{qx}\right)
        \left(g_{\nu \alpha} - \frac{1}{qx}(q_\nu x_\alpha + q_\alpha x_\nu) \right) q_\mu
        \right]
    \int {\cal D} \alpha_i e^{i (\alpha_{\bar q} + v \alpha_g) qx} {\cal T}_2(\alpha_i)
\nnb \\ &&+
        \frac{1}{qx} (q_\mu x_\nu - q_\nu x_\mu)
        (\varepsilon_\alpha q_\beta - \varepsilon_\beta q_\alpha)
    \int {\cal D} \alpha_i e^{i (\alpha_{\bar q} + v \alpha_g) qx} {\cal T}_3(\alpha_i)
\nnb \\ &&+
        \left. \frac{1}{qx} (q_\alpha x_\beta - q_\beta x_\alpha)
        (\varepsilon_\mu q_\nu - \varepsilon_\nu q_\mu)
    \int {\cal D} \alpha_i e^{i (\alpha_{\bar q} + v \alpha_g) qx} {\cal T}_4(\alpha_i)
                        \right\}, \nnb
\eea
where
$\varphi_\gamma(u)$ is the leading twist 2, $\psi^v(u)$,
$\psi^a(u)$, ${\cal A}$ and ${\cal V}$ are the twist 3 and
$h_\gamma(u)$, $\mathbb{A}$, ${\cal T}_i$ ($i=1,~2,~3,~4$) are the
twist 4 photon DAs, respectively and  $\chi$ is the magnetic susceptibility of the quarks.
The photon DAs is calculated in \cite{Rmag10}. The measure ${\cal D} \alpha_i$ is defined as
\bea
\int {\cal D} \alpha_i = \int_0^1 d \alpha_{\bar q} \int_0^1 d
\alpha_q \int_0^1 d \alpha_g \delta(1-\alpha_{\bar
q}-\alpha_q-\alpha_g).\nnb
\eea

Explicit form of the photon DAs entering into above matrix elements.

\bea
\varphi_\gamma(u) &=& 6 u \bar u \left( 1 + \varphi_2(\mu)
C_2^{\frac{3}{2}}(u - \bar u) \right),
\nnb \\
\psi^v(u) &=& 3 \left(3 (2 u - 1)^2 -1 \right)+\frac{3}{64} \left(15
w^V_\gamma - 5 w^A_\gamma\right)
                        \left(3 - 30 (2 u - 1)^2 + 35 (2 u -1)^4
                        \right),
\nnb \\
\psi^a(u) &=& \left(1- (2 u -1)^2\right)\left(5 (2 u -1)^2 -1\right)
\frac{5}{2}
    \left(1 + \frac{9}{16} w^V_\gamma - \frac{3}{16} w^A_\gamma
    \right),
\nnb \\
{\cal A}(\alpha_i) &=& 360 \alpha_q \alpha_{\bar q} \alpha_g^2
        \left(1 + w^A_\gamma \frac{1}{2} (7 \alpha_g - 3)\right),
\nnb \\
{\cal V}(\alpha_i) &=& 540 w^V_\gamma (\alpha_q - \alpha_{\bar q})
\alpha_q \alpha_{\bar q}
                \alpha_g^2,
\nnb \\
h_\gamma(u) &=& - 10 \left(1 + 2 \kappa^+\right) C_2^{\frac{1}{2}}(u
- \bar u),
\nnb \\
\mathbb{A}(u) &=& 40 u^2 \bar u^2 \left(3 \kappa - \kappa^+
+1\right) \nnb \\ && +
        8 (\zeta_2^+ - 3 \zeta_2) \left[u \bar u (2 + 13 u \bar u) \right.
\nnb \\ && + \left.
                2 u^3 (10 -15 u + 6 u^2) \ln(u) + 2 \bar u^3 (10 - 15 \bar u + 6 \bar u^2)
        \ln(\bar u) \right],
\nnb \\
{\cal T}_1(\alpha_i) &=& -120 (3 \zeta_2 + \zeta_2^+)(\alpha_{\bar
q} - \alpha_q)
        \alpha_{\bar q} \alpha_q \alpha_g,
\nnb \\
{\cal T}_2(\alpha_i) &=& 30 \alpha_g^2 (\alpha_{\bar q} - \alpha_q)
    \left((\kappa - \kappa^+) + (\zeta_1 - \zeta_1^+)(1 - 2\alpha_g) +
    \zeta_2 (3 - 4 \alpha_g)\right),
\nnb \\
{\cal T}_3(\alpha_i) &=& - 120 (3 \zeta_2 - \zeta_2^+)(\alpha_{\bar
q} -\alpha_q)
        \alpha_{\bar q} \alpha_q \alpha_g,
\nnb \\
{\cal T}_4(\alpha_i) &=& 30 \alpha_g^2 (\alpha_{\bar q} - \alpha_q)
    \left((\kappa + \kappa^+) + (\zeta_1 + \zeta_1^+)(1 - 2\alpha_g) +
    \zeta_2 (3 - 4 \alpha_g)\right),\nnb \\
{\cal S}(\alpha_i) &=& 30\alpha_g^2\{(\kappa +
\kappa^+)(1-\alpha_g)+(\zeta_1 + \zeta_1^+)(1 - \alpha_g)(1 -
2\alpha_g)\nnb \\&+&\zeta_2
[3 (\alpha_{\bar q} - \alpha_q)^2-\alpha_g(1 - \alpha_g)]\},\nnb \\
\tilde {\cal S}(\alpha_i) &=&-30\alpha_g^2\{(\kappa -
\kappa^+)(1-\alpha_g)+(\zeta_1 - \zeta_1^+)(1 - \alpha_g)(1 -
2\alpha_g)\nnb \\&+&\zeta_2 [3 (\alpha_{\bar q} -
\alpha_q)^2-\alpha_g(1 - \alpha_g)]\}.\nnb
\eea
The constants entering  the above DAs are obtained as
\cite{Rmag10} $\varphi_2(1~GeV) = 0$, $w^V_\gamma = 3.8 \pm 1.8$,
$w^A_\gamma = -2.1 \pm 1.0$, $\kappa = 0.2$, $\kappa^+ = 0$,
$\zeta_1 = 0.4$, $\zeta_2 = 0.3$, $\zeta_1^+ = 0$ and $\zeta_2^+ =
0$.

\section*{Appendix B}  
\setcounter{equation}{0}
\setcounter{section}{0}

In this Appendix we present the explicit expressions of the functions
$\Pi_i(u,d,s)$ for the magnetic moment of the $\Sigma^{0\ast}$ baryon
entering into the sum rule.

%
%
%
%
\bea
&&e^{m_{\Sigma_0}^2/M^2} \Pi_1 (u,d,s) = \nnb \\
&&{e \over 48} \Big[ (e_d-e_s) \dd \sp (1-t^2) -
(e_u+e_d) \uu \dd (1-t)^2 + (e_u-e_s) \uu \sp (1-t^2) \Big] \nnb \\
\cp \Big[i_2({\cal T}_3,1) - 2 i_2({\cal T}_3,v) - i_2({\cal T}_4,1) +
2 i_2({\cal T}_4,v) \Big] \nnb \\
\ar{e \over 192} f_{3\gamma} \Big\{ m_d \dd \Big[ e_s (1 + 6 t + t^2) -
e_u (1+t)^2 \Big] + m_s \sp (e_u+e_d) (1 + 6 t + t^2)\nnb \\
\ek m_u \uu \Big[ e_d (1+t)^2 - e_s (1 + 6 t + t^2)\Big]
\Big\} \Big[ i_3({\cal A},1) - 2  i_3({\cal A},v)\Big] \nnb \\
\ek {e \over 48} \Big[ 3 (e_d+e_s) \dd \sp (1-t^2)  + (e_u+e_d) \uu \dd 
(1-t)^2 + 3 (e_u+e_s) \uu \sp (1-t^2) \Big] i_2({\cal S},1) \nnb \\
\ek {e \over 192} f_{3\gamma} \Big\{ m_d \dd \Big[ e_s (3 + 2 t + 3
t^2) - e_u (1+t)^2 \Big] + m_s \sp (e_u+e_d) (3 + 2 t + 3 t^2)\nnb \\
\ek m_u \uu \Big[ e_d (1+t)^2 - e_s (3 + 2 t + 3 t^2)\Big]
\Big\} i_3({\cal V},1) \nnb \\
\ar {e \over 96} f_{3\gamma} \Big\{ \dd \Big[ - e_s m_d (1+t)^2 + 
e_u m_d (3 + 2 t + 3 t^2) + 6 e_u m_s (1-t^2) + 2 e_s m_u
(1-t)^2\Big] \nnb \\
\ar \sp \Big[6 (e_u m_d + e_d m_u) (1-t^2) + m_s (e_d + e_u) 
(3 + 2 t + 3 t^2) \Big] \nnb \\
\ar \uu \Big[ 2 e_s m_d (1-t)^2 + 6 e_d m_s (1-t^2) +
e_d m_u (3 + 2 t + 3 t^2) - e_s m_u (1+t)^2 \Big] \Big\}
\psi^a(u_0) \nnb \\
\ek {e \over 96} (1-t) \Big\{ \uu \sp (e_u-e_s) (1+t)
+ \dd \Big[ (e_d - e_s) \sp (1+t) - (e_u+e_d) \uu (1-t) \Big]
\Big\} \mathbb{A}(u_0) \nnb \\
\ek {e \over 1152 \pi^2} (1-t) \Big\{ e_s (m_u+m_d) \sp \gGgG -
\uu \Big[e_u \gGgG \Big( - m_d (1-t) + m_s  (1+t) \Big) \nnb \\
\ek 2 \pi^2 (e_u - 7 e_s) m_0^2 \sp (1+t) \Big]
+ \dd \Big[ - e_d \gGgG \Big( m_s (1+t) - m_u (1-t)\Big) \nnb \\
\ar 2 \pi^2 (e_d-7 e_s) m_0^2 \sp (1+t) -
6 \pi^2 (e_u+e_d) m_0^2 \uu (1-t)\Big\} \chi
\varphi_\gamma(u_0) \nnb \\
\ek {e \over 768 \pi^2} \Big\{ m_0^2 \sp \Big[ -18 (e_u m_d+e_d m_u)
(1-t^2) + m_s (e_u +e_d) (5+2 t + 5 t^2) \Big] \nnb \\
\ar \uu \Big[m_0^2 \Big( -6 e_s m_d (1-t)^2 - 18 e_d m_s (1-t^2) +
(e_d+e_s) m_u (5 + 2 t + 5 t^2) \Big) - 96 \pi^2 e_d \sp (1-t^2) 
\Big] \nnb \\
\ar \dd \Big[ e_u m_0^2 \Big( m_d (5 + 2 t + 5 t^2) - 18
m_s (1-t^2) \Big) +
e_s m_0^2 \Big( m_d (5 + 2 t + 5 t^2) - 6   
m_u (1-t)^2 \Big) \Big] \nnb \\
\ek 96 \pi^2 e_u \sp (1-t^2) - 32 \pi^2 e_s \uu (1-t)^2 \Big\} \nnb \\
%
%
\ar {e \over 768 \pi^2} \Bigg( \gamma_E - \ln{\Lambda^2\over M^2} \Bigg)    
\Big\{ (1-t)^2 \Big[ 3 m_0^2 (e_u m_u \dd + e_d m_d \uu) + \gGgG (e_u m_d
\uu + e_d m_u \dd) \chi \varphi_\gamma(u_0) \Big] \nnb \\
\ar (1-t^2) m_0^2 \Big[(18 e_u+7 e_s)  m_s \dd + (18 e_d+7 e_s) m_s \uu            
+ e_d (18 m_u - m_d) \sp  - e_d (m_u - 18 m_d) \sp \Big] \nnb \\
\ar (1-t^2) \gGgG \Big[ e_s (m_u + m_d) \sp - m_s (e_u \uu + e_d \dd) \Big] 
\chi \varphi_\gamma(u_0) \Big\} \nnb \\
%
%
\ek{e \over 64 \pi^2} M^2 \dd \Big\{ - e_s \Big[ m_d (1+t)^2 + 2
m_s(1-t^2) 
\Big] + 2 e_s m_u (1-t)^2 \nnb \\
\ar e_u \Big[ m_d (3 + 2 t + 3 t^2) + 6 
m_s (1-t^2) + 2 m_u (1-t)^2 \Big] \Big\} \nnb \\
\ek {e \over 64 \pi^2} M^2 \sp \Big\{ - e_d \Big[ 2 m_d (1-t^2)              
- m_s (3 + 2 t + 3 t^2) - 6 m_u (1-t^2) \Big] \nnb \\
\ar e_u \Big[ 6 m_d (1-t^2)
+ m_s (3 + 2 t + 3 t^2) - 2 m_u (1-t^2) \Big] \Big\} \nnb \\
\ek {e \over 64 \pi^2} M^2 \uu \Big\{ 2 (e_d + e_s) m_d (1-t)^2 +
2 (3 e_d + e_s) m_s (1-t^2) + m_u \Big[ e_d (3 + 2 t + 3 t^2)
- e_s (1+t)^2 \Big] \Big\} \nnb \\
\ar {e \over 128 \pi^2} M^2 (1-t) \Big\{ e_d \dd \Big[
-m_u (1-t) + m_s (1+t) \Big] -
e_s (m_u+m_d) \sp (1+t) \nnb \\
\ek e_u \uu \Big[
m_d (1-t) - m_s (1+t) \Big] \Big\} \mathbb{A}(u_0) \nnb \\
\ar {e \over 32 \pi^2} M^2 (1-t) \Big\{ e_d \dd \Big[m_u (1-t) + 2 m_s
(1+t)  \Big] + e_s (m_u+m_d) \sp (1+t) \nnb \\
\ar e_u \uu \Big[m_d (1-t) + 2 m_s (1+t) 
\Big] \Big\} i_2({\cal S},1) \nnb \\
\ek {e \over 64 \pi^2} M^2 (1-t) \Big\{ e_d \dd \Big[- m_u (1-t)
+ m_s (1+t) \Big] - e_s (m_u+m_d) \sp (1+t) \nnb \\
\ar e_u \uu \Big[- m_d (1-t) + m_s (1+t)  
\Big] \Big\} \Big[i_2(\widetilde{\cal S},1) + i_2({\cal T}_3,1) -
2 i_2({\cal T}_3,v) - i_2({\cal T}_4,1) + 2  i_2({\cal T}_4,v)\Big] \nnb \\
\ar {e \over 24} M^2 (1-t) \Big\{ (e_d-e_s) \dd \sp (1+t) 
- \uu \Big[(e_u+e_d) \dd (1-t) - (e_u-e_s) \sp (1+t) \Big] \Big\}
\chi \varphi_\gamma(u_0) \nnb \\
\ar {e \over 64 \pi^2} (1-t) M^2 \Bigg( \gamma_E - \ln{\Lambda^2\over M^2} \Bigg)
\Big\{ - 2 \dd \Big[e_u m_u (1-t) + e_s m_s (1+t) \Big] \nnb \\
\ar 2 (e_u m_u+e_d m_d) \sp (1+t) - 2 \uu \Big[e_d m_d (1-t) + e_s m_s (1+t)
\Big] \nnb \\
\ar e_d \dd \Big[m_u (1-t) + m_s (1+t) \Big] - e_s (m_u+m_d) \sp
(1+t) + e_u \uu \Big[m_d (1-t) + m_s (1+t) \Big] i_2({\cal
S},1) \nnb \\
\ar e_d \dd \Big[m_u (1-t) - m_s (1+t)\Big] +
e_s (m_u+m_d) \sp (1+t) + e_u \uu \Big[m_d (1-t) - m_s (1+t)
\Big] i_2(\widetilde{\cal S},1) \Big\} \nnb \\
\ek {e \over 1536 \pi^2} {1\over M^2} (1-t) \gGgG \Big\{
e_d \dd \Big[ m_u (1-t) + 3 m_s (1+t)\Big] +               
3 e_s (m_u+m_d) \sp (1+t) \nnb \\
\ar e_u \uu \Big[ m_d (1-t) + 3 m_s (1+t)\Big] \Big\} i_2({\cal S},1)\nnb \\
\ar {e \over 1536 \pi^2} {1\over M^2} (1-t) \gGgG \Big\{
e_d \dd \Big[ - m_u (1-t) + m_s (1+t)\Big] -
e_s (m_u+m_d) \sp (1+t) \nnb \\
\ek e_u \uu \Big[ m_d (1-t) - m_s (1+t)\Big]
 \Big\} \Big[i_2({\cal T}_3,1) -2 i_2({\cal T}_3,v)- i_2({\cal T}_4,1) + 2
i_2({\cal T}_4,v) \Big] \nnb \\
\ar {e \over 64} {1\over M^2} f_{3\gamma} m_0^2 (1-t^2) \Big[
m_s (e_u \dd + e_d \uu) + (e_u m_d + e_d m_u) \sp \Big]
\psi^a(u_0) \nnb \\
\ar {e \over 3072 \pi^2} {1\over M^2} \gGgG (1-t)  
\Big\{ e_d \dd \Big[m_u (1-t) - m_s (1+t)] + 
e_s (m_u+m_d) \sp (1+t) \nnb \\
\ar e_u \uu \Big[ m_d (1-t) - m_s (1+t) \Big] \Big\}
\mathbb{A}(u_0) \nnb \\ 
\ar {e \over 768 \pi^2} {1\over M^2} \gGgG (1-t) \Big\{
\dd \Big[3 e_u m_s (1+t) + e_s m_u (1-t)\Big] +
3 \sp (e_u m_d + e_d m_u) (1+t) \nnb \\
\ar \uu \Big[3 e_d m_s (1+t) + e_s m_d (1-t)\Big] \Big\} \nnb \\
\ar {e \over 9216 \pi^2} {1\over M^4} \gGgG (1-t) \Big\{
\dd \Big[ 3 e_u m_s (1+t) + e_s m_u (1-t) \Big]
+ 3 (e_u m_d+e_d m_u) \sp (1+t) \nnb \\
\ar \uu \Big[3 e_d m_s (1+t) + e_s m_d (1-t) \Big] \Big\}
\Big[3 m_0^2 + 8 \pi^2 f_{3\gamma} \psi^a(u_0)\Big] \nnb \\
\ar {e \over 2304} {1\over M^6} f_{3\gamma} m_0^2\gGgG (1-t) \Big\{
\dd \Big[ 3 e_u m_s (1+t) + e_s m_u (1-t) \Big]
+ 3 (e_u m_d+e_d m_u) \sp (1+t) \nnb \\
\ar \uu \Big[3 e_d m_s (1+t) + e_s m_d (1-t) \Big] \Big\}
\psi^a(u_0) \nnb \\
\ar {e \over 128 \pi^2} M^4 \Big\{ f_{3\gamma} \Big[ 2 t ( e_u +
e_d + 3 e_s) + e_s (1+t^2) \Big] \Big[i_3({\cal A},1) -
2 i_3({\cal A},v)\Big] \nnb \\
\ek f_{3\gamma} \Big[ (e_u+e_d) (1+t^2) + e_s (3 + 2 t + 3 t^2)\Big]
i_3({\cal V},1)  - 2 (1-t) \Big[ e_d \dd \Big( -m_u (1-t) + m_s
(1+t) \Big) \nnb \\
\ek e_s (m_u+m_d) \sp (1+t) +
e_u \uu \Big( -m_d (1-t) + m_s (1+t) \Big) \Big] \chi \varphi_\gamma 
(u_0) \nnb \\
\ar f_{3\gamma} \Big[(e_u+e_d) (3 + 2 t + 3 t^2) - e_s
(1+t)^2 \Big] \psi^a(u_0) \Big\} \nnb \\
\ek {e \over 256 \pi^4} M^6 \Big[ ( e_u + e_d) (3 + 2t +3 t^2) 
- e_s (1+t)^2 \Big] \nnb \\ \nnb \\ \nnb \\
&&e^{m_{\Sigma_0}^2/M^2} \Pi_2 (u,d,s) = \nnb \\
&& {e \over 192} \Big\{ \uu \sp \Big[4 (e_u+e_s) m_d (3 + 2 t + 3 t^2) -
e_u m_s (1-t^2) + e_s m_u (1-t)^2 \Big] \nnb \\
\ar \dd \sp 
\Big[ 4 (e_d+e_s) m_u (3 + 2 t + 3 t^2)  - e_d m_s (1-t^2)
+ e_s m_d (1-t)^2 \Big] \nnb \\
\ar \uu \dd \Big[ (e_u m_d+e_d m_u) (1-t^2) - 4 (e_u+e_d) m_s
(1+t)^2\Big ] \Big\} i_2({\cal S},1) \nnb \\
\ek {e \over 192} \Big\{ \dd \sp \Big[ 4 m_u (e_d+e_s) (1 + 6 t + t^2)
+ e_s m_d (1-t)^2 + e_d m_s (1-t^2)\Big] \nnb \\
\ek \uu \dd \Big[ (e_u m_d+e_d m_u) (1-t^2) + 4 (e_u + e_d) m_s
(1+t)^2 \Big] \nnb \\
\ar \uu \sp \Big[4 m_d (e_u+e_s) (1 + 6 t + t^2)+
e_s m_u (1-t)^2 + e_u m_s (1-t^2)\Big]
\Big\} i_2(\widetilde{\cal S},1) \nnb \\
\ar {e \over 192} \Big\{ \uu \dd (e_u m_d +e_d m_u) (1-t^2)
- \uu \sp \Big[ e_u m_s (1-t^2) + e_s m_u (1-t)^2\Big] \nnb \\
\ek \dd \sp \Big[ e_s m_d (1-t)^2 + e_d m_s (1-t^2) \Big]
\Big\} \Big[i_2({\cal T}_2,1) - 2 i_2({\cal T}_2,v)\Big] \nnb \\
\ar {e \over 96} e_s (m_u \uu + m_d \dd) \sp (1-t)^2    
\Big[i_2({\cal T}_3,1) - 2 i_2({\cal T}_3,v)\Big] \nnb \\
\ar {e \over 192} \Big\{ - \uu \dd (e_u m_d + e_d m_u) (1-t^2)
+\uu \sp \Big[ e_u m_s (1-t^2) - e_s m_u (1-t)^2 \nnb \\ 
\ek \dd \sp \Big[ e_s m_d (1-t)^2 - e_d m_s (1-t^2) \Big]
\Big\} \Big[i_2({\cal T}_4,1) - 2 i_2({\cal T}_4,v)\Big] \nnb \\
\ek {e \over 3072 \pi^2} f_{3\gamma} \gGgG (1-t)\Big\{
e_u \Big[m_d (1+t) - m_s (1-t)\Big] + 
e_d \Big[m_u (1+t) - m_s (1-t)\Big] \nnb \\
\ek e_s (m_u+m_d) (1+t) \Big\}
\Big[i_3({\cal A},1) - 2 i_3({\cal A},v)\Big] \nnb \\
\ek {e \over 384} (1-t)\Big\{                            
\uu \sp \Big[e_u m_s (1+t) + e_s m_u (1-t) \Big] +
\dd \sp \Big[e_d m_s (1+t) + e_s m_d (1-t) \Big]\nnb \\
\ek \uu \dd (e_u m_d + e_d m_u) (1+t) \Big\}
\Big[i_3(\widetilde{\cal S},1)+i_3({\cal T}_2,1) - 2 i_3({\cal T}_2,v)\Big] \nnb \\
\ar {e \over 192} e_s (m_u \uu + m_d \dd) \sp (1-t)^2
\Big[i_3({\cal T}_3,1) - 2 i_3({\cal T}_3,v)\Big] \nnb \\
\ek {e \over 384} (1-t)\Big\{
\uu \sp \Big[e_u m_s (1+t) - e_s m_u (1-t) \Big] +
\dd \sp \Big[e_d m_s (1+t) - e_s m_d (1-t)
\Big]\nnb \\
\ek \uu \dd (e_u m_d + e_d m_u) (1+t) \Big\}
\Big[i_3({\cal S},1) - i_3({\cal T}_4,1) + 2 i_3({\cal T}_4,v)\Big] \nnb \\
\ar {e \over 3072 \pi^2} f_{3\gamma} \gGgG (1-t)\Big\{  
e_u \Big[m_d (1+t) + m_s (1-t)\Big] +
e_d \Big[m_u (1+t) + m_s (1-t)\Big] \nnb \\
\ek e_s (m_u+m_d) (1+t) \Big\}
i_3({\cal V},1) \nnb \\
\ar {e \over 96} \Big\{ \uu \sp \Big[ 2 e_u m_d (6 + 4 t + 6 t^2)               
+ 3 e_u m_s (1-t^2) - e_s m_u (1-t)^2 - 2 e_s m_d (1+t)^2 \Big]
\nnb \\
\ar \dd \sp \Big[ e_d m_u (6 + 4 t + 6 t^2)
+ 3 e_d m_s (1-t^2) - e_s m_d (1-t)^2 - 2 e_s m_u (1+t)^2 \Big]
\nnb \\
\ar \uu \dd \Big[ (e_u+e_d) m_s (6 + 4 t + 6 t^2)
+ 3 (e_u m_d + e_d m_u) (1-t^2)\Big] \Big\}                                         
\Big[\widetilde{j}_1(h_\gamma) - 2 \widetilde{j}_2(h_\gamma) \Big] \nnb \\
\ek {e \over 576} m_0^2 \Big\{ \uu \sp \Big[ 4 e_u m_d (5 + 4 t + 5
t^2) + e_s m_u (1-t)^2 + 2 e_s m_d (1+t)^2 \Big] \nnb \\
\ar \dd \sp \Big[ 4 e_d m_u (5 + 4 t + 5   
t^2) + e_s m_d (1-t)^2 + 2 e_s m_u (1+t)^2 \Big] \nnb \\
\ar \uu \dd \Big[ 4 m_s (e_u + e_d) m_u (5 + 4 t + 5 t^2)\Big]
\Big\} \chi \varphi_\gamma(u_0) \nnb \\
\ar {e \over 2304 \pi^2} f_{3\gamma} (1-t) \Big\{ m_0^2 \pi^2 (1+t)
\Big[(e_d-7 e_s) \uu + (e_u-7 e_s) \dd\Big] - 3 (e_u+e_d) m_0^2 \pi^2\sp (1+t)
\nnb \\  
\ar e_u \gGgG \Big[ m_d (1+t) - m_s (1-t)\Big]
+ e_d \gGgG \Big[ m_u (1+t) - m_s (1-t)\Big] \nnb \\
\ek e_s \gGgG (1+t) (m_u+m_d) \Big\} \psi^a(u_0) \nnb \\
\ek {e \over 9216 \pi^2} f_{3\gamma} (1-t) \Big\{ 2 m_0^2 \pi^2 (1+t)
\Big[(e_d-7 e_s) \uu + (e_u-7 e_s) \dd\Big] - 6 (e_u+e_d) m_0^2 \pi^2\sp (1+t)
\nnb \\  
\ek e_u \gGgG \Big[ m_d (1+t) - m_s (1-t)\Big]
- e_d \gGgG \Big[ m_u (1+t) - m_s (1-t)\Big] \nnb \\
\ar e_s \gGgG (1+t) (m_u+m_d) \Big\} \psi^{a\prime}(u_0) \nnb \\
\ek {e \over 384} \Big\{ 4 e_d \uu \sp \Big[4 m_d (3+2 t +3 t^2)  
- m_s (1-t^2) + m_u (1-t)^2 \Big] \nnb \\
\ar e_s m_u \uu \sp (1-t)^2
\Big[2 \mathbb{A}(u_0) - \mathbb{A}^\prime (u_0)\Big] \Big\} \nnb \\
\ek {e \over 384} \Big\{ 4 e_u \dd \sp \Big[4 m_u (3+2 t +3 t^2)
- m_s (1-t^2) + m_d (1-t)^2 \Big] \nnb \\
\ar e_s m_d \dd \sp (1-t)^2
\Big[2 \mathbb{A}(u_0) - \mathbb{A}^\prime (u_0)\Big] \Big\} \nnb \\
\ek {e \over 96} e_s \uu \dd (1+t) \Big\{
\Big[ (1-t) (m_u+m_d) - 4 m_s (1+t) \Big] \Big\} \nnb \\
\ar {e\over 1152} e_s m_0^2 \sp (1-t)^2 (m_u \uu + m_d \dd) \chi
\varphi_\gamma^\prime(u_0) \nnb \\
\ar {e\over 512 \pi^2} M^4 e_s\sp (1-t)^2 \Big\{
2 \mathbb{A}(u_0) + \mathbb{A}^\prime(u_0)-
2 \Big[6 i_2({\cal S},1) + 2 i_2(\widetilde{\cal
S},1) + 2 i_2({\cal T}_2,1) \nnb \\
\ar 4 i_2({\cal T}_3,1) - 6 i_2({\cal T}_4,1) -
4 i_2({\cal T}_2,v) - 8 i_2({\cal T}_3,v) + 12 i_2({\cal T}_4,v) -
i_3({\cal S},1) + i_3(\widetilde{\cal S},1) \nnb \\
\ar i_3({\cal T}_2,1) -
2 i_3({\cal T}_3,1) + i_3({\cal T}_4,1) - 2 
i_3({\cal T}_2,v) + 4 i_3({\cal T}_3,v) - 2 i_3({\cal T}_4,v)
+ 2 \widetilde{j}_1(h_\gamma) - 4 \widetilde{j}_2(h_\gamma) \Big]
\Big\} \nnb \\
\ar {e\over 512 \pi^2} M^4 f_{3\gamma} (1-t) \Big\{
e_u \Big[m_d (1+t) - m_s(1-t) \Big] +
e_d \Big[m_u (1+t) - m_s(1-t) \Big] \nnb \\
\ek e_s (m_u+m_d) (1+t) \Big\}
\Big[4 i_3({\cal A},1) - 8 i_3({\cal A},v) - 6 \psi^a(u_0) - \psi^{a\prime}(u_0) \Big] \nnb \\
\ar {3 e\over 128 \pi^2} M^4 (e_u \uu + e_d \dd)  (1-t^2) \Big[
\widetilde{j}_1(h_\gamma) - 2 \widetilde{j}_2(h_\gamma) \Big] \nnb \\
\ar {3 e\over 128 \pi^2} M^4(1-t) \Big\{
(1+t) \Big[ (e_u \uu + e_d \dd) \mathbb{A}(u_0)  
-(e_d -e_s) \uu - (e_u -e_s) \dd \Big] \nnb \\
\ar (e_u+e_d) \sp (1-t) \Big\} \nnb \\
\ek {e\over 128 \pi^2} M^4 f_{3\gamma} (1-t) \Big\{
e_u \Big[5 m_d (1+t) + m_s(1-t) \Big] +
e_d \Big[5 m_u (1+t) + m_s(1-t) \Big] \nnb \\
\ar e_s (m_u+m_d) (1+t) \Big\} i_3({\cal V},1) \nnb \\
\ar {e\over 64 \pi^2} M^4 f_{3\gamma} (1-t^2)
\Bigg( \gamma_E - \ln{\Lambda^2\over M^2} \Bigg)
\Big[2 (e_u m_d + e_d m_u) + e_s (m_u+m_d) \Big] i_3({\cal V},1) \nnb \\
\ar {e\over 6144 \pi^2} f_{3\gamma} \gGgG (1-t)
\Bigg( \gamma_E - \ln{\Lambda^2\over M^2} \Bigg)
\Big\{
e_u \Big[m_d (1+t) - m_s(1-t) \Big] \nnb \\
\ar e_d \Big[m_u (1+t) - m_s(1-t) \Big]
- e_s (m_u+m_d) (1+t) \Big\}
\Big[2 \psi^a(u_0) - \psi^{a\prime}(u_0)\Big] \nnb \\
\ek {e\over 2048 \pi^4} M^2 \gGgG (1-t)
\Bigg( \gamma_E - \ln{\Lambda^2\over M^2} \Bigg)
\Big\{   
e_u \Big[m_d (1+t) - m_s(1-t) \Big] +
e_d \Big[m_u (1+t) - m_s(1-t) \Big] \nnb \\
\ek e_s (m_u+m_d) (1+t) \Big\} \nnb \\
\ek {e\over 192} M^2 f_{3\gamma} (1-t) \Big\{
e_u \Big[\dd (1+t) - \sp (1-t) \Big] +
  e_d \Big[\uu  (1+t) - \sp (1-t) \Big] \nnb \\
\ek e_s (\uu+\dd) (1+t) \Big\}
\Big\{2 i_3({\cal A},1) - 2 \Big[i_3({\cal V},1) + 2 i_3({\cal A},v)
+ \psi^a(u_0) \Big] -
\psi^{a\prime}(u_0) \Big\} \nnb \\
\ar {e \over 6144 \pi^4} M^2 (1-t) \Big\{
  4 e_u m_0^2 \pi^2 \Big[ \dd (1+t) - 3 \sp (1-t)\Big]
- 5 e_u \gGgG \Big[ m_d (1+t) - m_s (1-t)\Big] \nnb \\
\ar 4 e_d m_0^2 \pi^2 \Big[ \uu (1+t) - 3 \sp (1-t)\Big]  
- 5 e_d \gGgG \Big[ m_u (1+t) - m_s (1-t)\Big] \nnb \\
\ek e_s (1+t) \Big[28 m_0^2 \pi^2 (\uu+\dd) - 5 \gGgG (m_u+m_d) \Big]
\Big\} \nnb \\
\ar {e \over 48} M^2 (1-t)^2 (e_u+e_d) f_{3\gamma} \sp i_3({\cal V},1)
\nnb \\
\ar {e \over 48} M^2 \Big\{ \uu \sp \Big[ 2 e_s m_d (1+t)^2 -
e_u m_d (6 + 4 t + 6 t^2)               
- 3 e_u m_s (1-t^2) \Big] \nnb \\
\ek \dd \sp \Big[ 3 e_d m_s (1-t^2) + e_d m_u (6 + 4 t + 6 t^2)
- 2 e_s m_u (1+t)^2 \Big] \nnb \\
\ek \uu \dd \Big[ 3 e_u m_d (1-t^2) + (e_u +e_d) m_s (6 + 4 t + 6 t^2)
+ 3 e_d m_u (1-t^2)\Big] \Big\}                                         
\chi \varphi_\gamma (u_0) \nnb \\
\ek {e \over 96} M^2 e_s (m_u \uu + m_d \dd) \sp (1-t)^2 \chi
\varphi_\gamma^\prime (u_0) \nnb \\
\ek {e \over 256\pi^4} M^6 
\Bigg( \gamma_E - \ln{\Lambda^2\over M^2} \Bigg) (1-t)
\Big[ 9 (e_u m_u + e_d m_d) (1+t) + 2 e_s m_s (1-t) \Big] \nnb \\
\ek {e \over 1536\pi^4} M^6 (1-t) \Big\{
3 e_u \Big[ (1+t) ( 27 m_u - 5 m_d) +  5 m_s (1-t) \Big] \nnb \\
\ar 3 e_d \Big[ (1+t) ( 27 m_d - 5 m_u) +  5 m_s (1-t) \Big] +
e_s \Big[ 15 (1+t) (m_u + m_d) +  16 m_s (1-t) \Big]
\Big\} \nnb \\
\ek {e \over 192 \pi^2} M^6 (1-t)
\Big[ 9 (e_u \uu + e_d \dd) (1+t) + 2 e_s \sp (1-t) \Big]
\chi \varphi_\gamma(u_0) \nnb \\
\ek {e \over 384 \pi^2} M^6 (1-t)^2 e_s \sp \chi \varphi_\gamma^\prime
(u_0) \nnb \\
\ek {e \over 1152} {1 \over M^2} m_0^2 \Big\{ 
\uu \sp \Big[ 2 e_u m_d (5 + 4 t + 5 t^2) + e_s m_d (1+t)^2 \Big] \nnb \\
\ar \dd \sp \Big[ 2 e_d m_u (5 + 4 t + 5 t^2) + e_s m_u (1+t)^2 \Big] \nnb \\
\ar 2 (e_u + e_d) m_s \uu \dd (5 + 4 t + 5 t^2) \Big\} \mathbb{A}(u_0) \nnb \\
\ar {e \over 192} {1 \over M^2} m_0^2 \Big\{ 
(3 + 2 t + 3 t^2) \Big[\uu \sp (e_u + e_s) m_d +
\dd \sp (e_d + e_s) m_u \Big] \nnb \\
\ek (e_u + e_d) m_s \uu \dd (1+t)^2 \Big\} i_2({\cal S},1) \nnb \\
\ek {e \over 192} {1 \over M^2} m_0^2 \Big\{ 
(1 + 6 t + t^2) \Big[\uu \sp (e_u + e_s) m_d +
\dd \sp (e_d + e_s) m_u \Big] \nnb \\
\ek (e_u + e_d) m_s \uu \dd (1+t)^2 \Big\} i_2(\widetilde{\cal S},1) \nnb \\
\ek {e \over 576} {1 \over M^2} \gGgG \Big\{
\uu \sp \Big[ e_s m_d (1+t)^2 - e_u m_d (3 + 2 t + 3 t^2) \Big] \nnb \\
\ek \dd \sp \Big[ e_s m_u (1+t)^2 - e_d m_u (3 + 2 t + 3 t^2) \Big] \nnb \\
\ar \uu \dd (e_u+e_d) m_s (3 + 2 t + 3 t^2) \Big\} \chi \varphi_\gamma(u_0) \nnb \\
\ek {e \over 576} {1 \over M^2} m_0^2 \Big\{
\uu \sp \Big[ 2 e_u m_d (5 + 4 t + 5 t^2) + e_s m_d (1+t)^2
\Big] \nnb \\
\ar \dd \sp \Big[ 2 e_d m_u (5 + 4 t + 5 t^2) + e_s m_u (1+t)^2
\Big] \nnb \\
\ar 2 (e_u + e_d) m_s \uu \dd (5 + 4 t + 5 t^2) \Big\}
\Big[\widetilde{j}_1(h_\gamma) - 2 \widetilde{j}_2(h_\gamma) \Big] \nnb \\
\ek {e \over 48} {1 \over M^2} m_0^2 \Big[
(e_u m_u \dd + e_d m_d \uu) \sp (3 + 2 t + 3 t^2) -
e_s m_s \uu \dd (1+t)^2 \Big] \nnb \\
\ar {e \over 1152} {1 \over M^4} \gGgG \Big\{
\uu \sp \Big[ e_s m_d (1+t)^2 - e_u m_d (3 + 2 t + 3 t^2) \Big] \nnb \\
\ar \dd \sp \Big[ e_s m_u (1+t)^2 - e_d m_u (3 + 2 t + 3 t^2) \Big] \nnb \\
\ek \uu \dd (e_u+e_d) m_s (3 + 2 t + 3 t^2) \Big\} \Big[
\mathbb{A}(u_0) + m_0^2 \chi \varphi_\gamma(u_0) - \widetilde{j}_1(h_\gamma)
+ 2 \widetilde{j}_2(h_\gamma) \Big] \nnb \\
\ar {e \over 4608} {1 \over M^6} \gGgG m_0^2 \Big\{
\uu \sp \Big[ e_s m_d (1+t)^2 - e_u m_d (3 + 2 t + 3 t^2) \Big] \nnb \\
\ar \dd \sp \Big[ e_s m_u (1+t)^2 - e_d m_u (3 + 2 t + 3 t^2) \Big] \nnb \\
\ek \uu \dd (e_u+e_d) m_s (3 + 2 t + 3 t^2) \Big\} \Big[
3 \mathbb{A}(u_0) - 2 \widetilde{j}_1(h_\gamma)
+ 4 \widetilde{j}_2(h_\gamma) \Big] \nnb \\ \nnb \\ \nnb \\
&&e^{m_{\Sigma_0}^2/M^2} \Pi_3 (u,d,s) = \nnb \\
&& - {e \over 6144 \pi^2} \gGgG (1-t) \Big\{ 
e_u \uu \Big[ m_d (1-t) - m_s (1+t) \Big] +
e_d \dd \Big[ m_u (1-t) - m_s (1+t) \Big] \nnb \\
\ar e_s (m_u+m_d) \sp (1+t) \Big\}
\Big\{ 4 i_3({\cal S},1) + 2 \Big[i_3({\cal T}_1,1) + i_3({\cal T}_3,1) \nnb \\
\ek 2 i_3({\cal T}_4,1) - 2 i_3({\cal S},v) -
2 i_3({\cal T}_3,v) + 2 i_3({\cal T}_4,v) +
\mathbb{A}^\prime(u_0)\Big\} \nnb \\
\ar {e \over 128} f_{3\gamma} m_0^2 (1-t^2)
\Big[ e_u (m_d \sp + m_s \dd) + e_d (m_u \sp + m_s \uu) \Big]
\Big[4 \psi^v(u_0) - \psi^{a\prime} (u_0) \Big] \nnb \\
\ek {e \over 1536 \pi^2} \gGgG (1-t^2) \Big\{ 
\Big[ m_s (e_u \uu + e_d \dd) - e_s (m_u + m_d) \sp \Big]
\Big[i_3({\cal S},1) + i_3({\cal T}_1,1) \nnb \\
\ek i_3({\cal T}_4,1) \Big]
+ 2 \Big[ m_s (e_u \uu + e_d \dd) + 2 e_s (m_u + m_d) \sp \Big]
i_3({\cal S},v) \Big\} \nnb \\
\ek {e \over 2304 \pi^2} \gGgG (1-t) \Big\{
\uu \Big[ 3 e_d m_s (1+t) - e_s m_d (1-t) \Big] +
\dd \Big[ 3 e_u m_s (1+t) - e_s m_u (1-t) \Big] \nnb \\
\ar 3 (e_u m_d + e_d m_u) \sp  (1+t) \Big\} \nnb \\
\ek {e \over 48} m_0^2 (1-t) \Big[
3 (1+t) (e_u \dd + e_d \uu) \sp - (1-t) e_s \uu \dd \Big] \nnb \\
\ek {e\over 256 \pi2} M^4 (1-t) \Big\{
e_u \uu \Big[m_d (1-t) - m_s (1+t)\Big] +
e_d \dd \Big[m_u (1-t) - m_s (1+t)\Big] \nnb \\
\ar e_s \sp (m_u+m_d) \sp (1+t) \Big\}
\Big\{2 i_3({\cal S},1) - 2 \Big[i_3(\widetilde{\cal S},1) - 
2 i_3({\cal T}_1,1) + i_3({\cal T}_2,1) + i_3({\cal T}_4,1) \nnb \\
\ek 6 i_3({\cal S},v) - 2 i_3({\cal T}_3,v) + 2 i_3({\cal T}_4,v) \Big]
- \mathbb{A}^\prime(u_0) \Big\} \nnb \\
\ar {e\over 32 \pi^2} M^4 (1-t)^2 (e_u m_d \uu +
e_d m_u \dd) \Big[i_3({\cal S},1) + i_3({\cal T}_1,1) -
i_3({\cal T}_4,1) + i_3({\cal S},v) \Big] \nnb \\
\ar {3 e\over 32 \pi^2} M^4 (1-t^2) e_s (m_u+m_d) \sp
i_3({\cal S},v) \nnb \\
\ek {e\over 48} M^4 (1-t) \Big\{
\Big[(e_u - e_s) \uu \sp + (e_d-e_s) \dd \sp \Big] (1+t) 
- (e_u + e_d) \uu \dd (1-t) \Big\}
\chi \varphi_\gamma^\prime (u_0) \nnb \\
\ar {e\over 128 \pi^2} M^4 \uu \Big\{
2 (e_d - e_s) (1-t) \Big[ m_d (1-t) + 3 m_s (1+t) \Big]   
+ m_u \Big[ e_d (1+t)^2 - e_s (7+6 t + 7 t^2) \Big] \Big\} \nnb \\
\ar {e\over 128 \pi^2} M^4 \dd \Big\{
e_u \Big[ -  2 m_u (1-t) + m_d (1+t) + 6 m_s (1-t) \Big] (1+t)  
- e_s m_d (7+6 t + 7 t^2) \nnb \\
\ek 2 e_s \Big[ m_u(1-t) + 3 m_s (1+t)\Big] (1-t) \Big\} \nnb \\
\ek {e\over 128 \pi^2} M^4 \sp (1+t) \Big\{                    
e_u \Big[ 6 (m_u-m_d) (1-t) - m_s (1+t) \Big]
- e_d \Big[ 6 (m_u-m_d) (1-t) + m_s (1+t) \Big] \Big\} \nnb \\
\ar {e\over 128 \pi^2} M^4 \Bigg( \gamma_E - \ln{\Lambda^2\over M^2} \Bigg)
(1-t) \Big\{ e_u \uu \Big[m_d (1-t) + m_s (1+t) \Big] \nnb \\
\ar e_d \dd \Big[m_u (1-t) + m_s (1+t) \Big] +
e_s \sp (1+t) (m_u+m_d) \Big\} \nnb \\
\cp \Big[i_3({\cal S},1) - i_3(\widetilde{\cal S},1) + i_3({\cal T}_1,1) -
i_3({\cal T}_2,1) + i_3({\cal T}_3,1) - i_3({\cal T}_4,1) \Big] \nnb \\
\ar {e\over 64 \pi^2} M^4 \Bigg( \gamma_E - \ln{\Lambda^2\over M^2} \Bigg)
(1-t)^2 (e_u m_d \uu + e_d m_u \dd) \Big[i_3(\widetilde{\cal S},1) +
i_3({\cal T}_2,1) - i_3({\cal T}_3,1) \Big] \nnb \\
\ek {e\over 768 \pi^2} \gGgG \Bigg( \gamma_E - \ln{\Lambda^2\over M^2} \Bigg)
(1-t) \Big\{ 3 (1+t) \Big[ e_u (m_d \sp + m_s \dd) + e_d (m_u \sp +
m_s \uu) \Big] \nnb \\
\ek (1-t) e_s (m_u \dd + m_d \uu) \Big\} \nnb \\
\ek {e\over 1536 \pi^2} M^2 \gGgG \Bigg( \gamma_E - \ln{\Lambda^2\over M^2}
\Bigg) (1-t) \Big\{
e_u \uu \Big[ m_d (1-t) - m_s (1+t) \Big] \nnb \\
\ar e_d \dd \Big[ m_u (1-t) - m_s (1+t) \Big]
+ e_s \sp (m_u + m_d) (1+t) \Big\}
\chi \varphi_\gamma^\prime(u_0) \nnb \\
\ar {e\over 96} M^2 (1-t) \Big\{
(1+t) \Big[(e_u-e_s) \uu + (e_d-e_s) \dd \Big] \sp \nnb \\
\ek (e_u + e_d) \uu \dd  (1-t) \Big\}
\Big[ i_3({\cal T}_3,1) - 2 i_3({\cal T}_3,v) + 
2 i_3({\cal T}_4,v) \Big] \nnb \\
\ek {e\over 384} M^2 (e_u+e_d) \uu \dd (1-t)^2 
\Big[8 i_3({\cal S},1) + 4 i_3({\cal T}_1,1) - 8 i_3({\cal T}_4,1)
-8  i_3({\cal S},v) \nnb \\
\ar 2 \mathbb{A}^\prime (u_0) +
m_0^2 \chi \varphi_\gamma^\prime(u_0) \Big] \nnb \\
\ek {e\over 96} M^2 (1-t^2) \sp \Big\{
\Big[ (e_u-e_s) \uu + (e_d-e_s) \dd \Big] i_3({\cal T}_1,1) \nnb \\
\ar 6 \Big[ (e_u+e_s) \uu + (e_d+e_s) \dd \Big] i_3({\cal S},v)
\Big\} \nnb \\
\ar {e\over 192} M^2 f_{3\gamma} \Big\{    
m_u \uu \Big[e_d (1+t)^2 - e_s (1 + 6 t + t^2)\Big]
+ m_d \dd \Big[e_u (1+t)^2 - e_s (1 + 6 t + t^2)\Big] \nnb \\
\ek (e_u + e_d) m_s \sp (1 + 6 t + t^2) \Big\}
i_4({\cal A},v) \nnb \\
\ar {e\over 192} M^2 f_{3\gamma} \Big\{
m_u \uu \Big[e_d (1+t)^2 - e_s (3 + 2 t + 3 t^2)\Big]
+ m_d \dd \Big[e_u (1+t)^2 - e_s (3 + 2 t + 3 t^2)\Big] \nnb \\
\ek (e_u + e_d) m_s \sp (3 + 2 t + 3 t^2) \Big\}
i_4({\cal V},v) \nnb \\
\ar {e\over 192} M^2 (1-t^2) \Big[(e_u-e_s) \uu + (e_d-e_s) \dd \Big]\sp
\mathbb{A}^\prime (u_0) \nnb \\
\ar {e\over 48} M^2 f_{3\gamma} \uu \Big\{
e_d \Big[  m_u (3 + 2 t + 3 t^2) + 6 m_s (1+t^2)\Big] - 
e_s \Big[ m_u (1+t)^2 + 2 m_d  (1-t)^2\Big]\Big\}
\psi^v(u_0) \nnb \\
\ar {e\over 48} M^2 f_{3\gamma} \dd \Big\{
e_u \Big[ m_d (3 + 2 t + 3 t^2) + 6 m_s(1-t^2)\Big] -
e_s \Big[ 2 m_u (1-t)^2 + m_d  (1+t)^2\Big]\Big\}
\psi^v(u_0) \nnb \\
\ar {e\over 48} M^2 f_{3\gamma} \sp \Big\{
e_u \Big[ m_s (3 + 2 t + 3 t^2) + 6 m_d(1-t^2)\Big] +
e_d \Big[ m_u (1-t)^2 + m_s (3 + 2 t + 3 t^2)\Big]\Big\}
\psi^v(u_0) \nnb \\
\ar {e\over 2304 \pi^2} M^2 e_u \gGgG \uu (1-t)
\Big[m_d (1-t) - m_s (1+t) \Big]
\chi \varphi_\gamma^\prime(u_0) \nnb \\
\ar {e\over 1152} M^2 m_0^2 (e_u -7 e_s) \uu \sp (1-t^2)
\chi \varphi_\gamma^\prime(u_0) \nnb \\
\ar {e\over 2304 \pi^2} M^2 e_d \gGgG \dd (1-t)
\Big[m_u (1-t) - m_s (1+t) \Big]
\chi \varphi_\gamma^\prime(u_0) \nnb \\
\ar {e\over 1152} M^2 m_0^2 (e_d -7 e_s) \dd \sp (1-t^2)
\chi \varphi_\gamma^\prime(u_0) \nnb \\
\ar {e\over 2304 \pi^2} M^2 e_s (m_u+m_d) \gGgG \sp (1-t^2) \chi
\varphi_\gamma^\prime(u_0) \nnb \\
\ek {e\over 192} M^2 f_{3\gamma} \uu \Big\{
e_d \Big[ m_u (3 + 2 t + 3 t^2) + 6 m_s (1-t^2)\Big] + 
e_s \Big[ m_u (1+t)^2 + 2 m_d  (1-t)^2\Big]\Big\}
\psi^{a\prime}(u_0) \nnb \\
\ek {e\over 192} M^2 f_{3\gamma} \dd \Big\{
e_u \Big[ m_d (3 + 2 t + 3 t^2) + 6 m_s(1-t^2)\Big]
+ e_s \Big[ 2 m_u (1-t)^2 - m_d  (1+t)^2\Big]\Big\}
\psi^{a\prime}(u_0) \nnb \\
\ek {e\over 192} M^2 f_{3\gamma} \sp \Big\{
e_u \Big[ m_s (3 + 2 t + 3 t^2) + 6 m_d (1-t^2)\Big] +
e_d \Big[ m_u (1-t)^2 + m_s (3 + 2 t + 3 t^2)\Big]\Big\}
\psi^{a\prime}(u_0) \nnb \\
\ek {e\over 768 \pi^2 } M^2 m_0^2 \uu \Big\{
e_d \Big[ 3 m_d (1-t)^2 - 18 m_s (1-t^2) -   
4 m_u (+t+t^2) \Big] \nnb \\
\ek e_s \Big[ 21 m_s (1-t^2) + 2 m_u (7+10 t+7t^2) \Big\} \nnb \\
\ar {e\over 768 \pi^2 } M^2 m_0^2 \dd \Big\{
e_u \Big[4 m_d (1+t+t^2) + 18 m_s (1-t^2) -
3 m_u (1-t)^2 \Big] \nnb \\
\ar e_s \Big[ 2 m_d (7+10 t+7t^2) + 21 m_s (1-t^2)\Big] \Big\} \nnb \\ 
\ar {e\over 768 \pi^2 } M^2 m_0^2 \sp \Big\{
e_u \Big[18 m_d (1-t^2) + 4 m_s (1+t+t^2) +
15 m_u (1-t)^2 \Big] \nnb \\
\ar e_d \Big[15 m_d (1-t^2) + 4 m_s (1+t+t^2) + 18 m_u
(1-t^2)\Big] \Big\} \nnb \\
\ar {e\over 1024 \pi^4 } M^8 \Big[
(e_u +e_d) (7+6 t+7t^2) - e_s (9+10t+9t^2)\Big] \nnb \\
\ek {e\over 128 \pi^2 } M^6 f_{3\gamma}\Big[2 (e_u+e_d) +
e_s (1+6t+t^2)\Big] i_4({\cal A},v) \nnb \\
\ek {e\over 128 \pi^2 } M^6 f_{3\gamma}\Big[(e_u+e_d) (1+t^2) +
e_s (3+2t+3t^2)\Big] i_4({\cal V},v) \nnb \\
\ar {e\over 64 \pi^2 } M^6 f_{3\gamma}\Big[(e_u+e_d) (3+2t+3t^2)
- e_s ((1+t)^2)\Big] \psi^v(u_0) \nnb \\
\ek {e\over 128 \pi^2 } M^6 (1-t) \Big\{
e_u \uu \Big[m_d (1-t) - m_s (1+t)\Big] +
e_d \dd \Big[m_u (1-t) - m_s (1+t)\Big] \nnb \\
\ar e_s \sp (m_u+ m_d) (1+t)\Big\} \chi\varphi_\gamma^\prime (u_0) \nnb \\
\ek {e\over 256 \pi^2 } M^6 f_{3\gamma}\Big[(e_u+e_d) (3+2t+3t^2) 
- e_s ((1+t)^2)\Big] \psi^{a\prime}(u_0) \nnb \\
\ek {e\over 4608 \pi^2 } {1\over M^2} \gGgG (1-t) \Big\{
\uu \Big[ 3 e_d m_s (1+t)- e_s m_d (1-t)\Big] \nnb \\
\ar \dd \Big[ 3 e_u m_s (1+t)- e_s m_u (1-t)\Big] +
3 \sp (e_u m_d + e_d m_u)  (1+t) \Big\}
\Big[ 3 m_0^2 - 8 \pi^2 f_{3\gamma} \psi^v (u_0)\Big] \nnb \\
\ek {e\over 2304} {1\over M^2} f_{3\gamma} \gGgG (1-t) \Big\{
\uu \Big[ 3 e_d m_s (1+t)+ e_s m_d (1-t)\Big] \nnb \\
\ar \dd \Big[ 3 e_u m_s (1+t)+ e_s m_u (1-t)\Big]
+ 3 \sp (e_u m_d + e_d m_u)  (1+t) \Big\}
\psi^{a\prime} (u_0)\nnb \\
\ar {e\over 1152} {1\over M^4} m_0^2 f_{3\gamma} \gGgG (1-t) \Big\{
\uu \Big[ 3 e_d m_s (1+t)- e_s m_d (1-t)\Big] \nnb \\
\ar \dd \Big[ 3 e_u m_s (1+t)- e_s m_u (1-t)\Big]
+ 3 \sp (e_u m_d + e_d m_u)  (1+t) \Big\}
\Big[ 3 m_0^2 - 8 \pi^2 f_{3\gamma} \psi^v (u_0) \Big] \nnb \\
\ek {e\over 4608} {1\over M^4} m_0^2 f_{3\gamma} \gGgG (1-t)\Big\{
\uu \Big[ 3 e_d m_s (1+t)+ e_s m_d (1-t)\Big] \nnb \\
\ar \dd \Big[ 3 e_u m_s (1+t)+ e_s m_u (1-t)\Big]
+ 3 \sp (e_u m_d + e_d m_u)  (1+t) \Big\}
\Big[ 3 m_0^2 - 8 \pi^2 f_{3\gamma} \psi^{a\prime} (u_0) \Big] \nnb \\ \nnb \\ \nnb \\
&&e^{m_{\Sigma_0}^2/M^2} \Pi_4 (u,d,s) = \nnb \\
&& {e\over 192} (1-t) \Big\{ \uu\sp \Big[e_u m_s (1+t) - e_s m_u
(1-t) \Big] + \dd \sp \Big[e_d m_s (1+t) - e_s m_d
(1-t) \Big] \nnb \\
\ek \uu \dd (e_u m_d + e_d m_u)(1-t^2) 
\Big\} \Big[i_3({\cal S},1) - i_3({\cal T}_4,1) +
2 i_3({\cal T}_4,v)\Big] \nnb \\
\ar {e\over 192} (1-t) \Big\{ \uu\sp \Big[e_u m_s (1+t) + e_s m_u
(1-t) \Big] + \dd \sp \Big[e_d m_s (1+t) + e_s m_d
(1-t) \Big] \nnb \\
\ek \uu \dd (e_u m_d + e_d m_u)(1-t^2)         
\Big\} \Big[i_3(\widetilde{\cal S},1) + i_3({\cal T}_2,1)\Big] \nnb \\
\ek {e\over 576} e_s (m_u \uu + m_d \dd) \sp (1-t)^2 \Big\{
6 \Big[i_3({\cal T}_3,1) + i_3({\cal T}_2,v) - 2 i_3({\cal T}_3,v)
- 2 \widetilde{j}_1(h_\gamma) \Big] \nnb \\
\ar 3 \mathbb{A}^\prime (u_o) + m_0^2 \chi
\varphi_\gamma^\prime (u_0) \Big\} \nnb \\
\ek {e\over 96} (1-t^2) \Big[ 
e_u m_s \uu \sp + e_d m_s \dd \sp -
(e_u m_d + e_d m_u) \uu \dd \Big] i_3({\cal T}_2,v) \nnb \\
\ek {e\over 48} \Big\{ \dd \sp \Big[3 e_d m_s (1-t^2) +
2 e_d m_u (3 + 2 t + 3 t^2) - 2 e_s m_u (1+t)^2 \Big] \nnb \\
\ar \uu \sp \Big[3 e_u m_s (1-t^2) +
2 e_u m_d (3 + 2 t + 3 t^2) - 2 e_s m_d (1+t)^2 \Big] \nnb \\
\ar \uu \dd \Big[3 (e_u m_d + e_d m_u) (1-t^2) +
2 (e_u + e_d ) m_s (3 + 2 t + 3 t^2) \Big]
\Big\} \widetilde{j}_1(h_\gamma) \nnb \\
\ar {e\over 2304} m_0^2 f_{3\gamma} (1-t) \Big\{
(1+t) \Big[\uu (e_d - 7 e_s) +
\dd (e_u - 7 e_s) \Big] -
3 \sp (e_u+e_d) (1-t) \Big\} \psi^{a\prime}(u_0) \nnb \\
\ek {e\over 4608 \pi^2} \gGgG f_{3\gamma} (1-t) \Big\{ 
e_u \Big[m_d (1+t) - m_s (1-t) \Big]+
e_d \Big[m_u (1+t) - m_s (1-t) \Big] \nnb \\
\ek e_s (m_u+m_d) (1+t) \Big\} \psi^{a\prime}(u_0) \nnb \\
\ar {e\over 3072 \pi^2} \gGgG f_{3\gamma} (1-t) 
\Bigg( \gamma_E - \ln{\Lambda^2\over M^2} \Bigg)
\Big\{ e_u \Big[m_d (1+t) - m_s (1-t) \Big] \nnb \\
\ar e_d \Big[m_u (1+t) - m_s (1-t) \Big] -
e_s (m_u+m_d) (1+t) \Big\} \psi^{a\prime}(u_0) \nnb \\
\ar {e\over 48} M^2 e_s (m_u \uu + m_d \dd) \sp (1-t)^2 
\chi \varphi_\gamma^\prime (u_0) \nnb \\
\ek {e\over 96} M^2 f_{3\gamma} (1-t)
\Big\{ e_u \Big[\dd (1+t) - \sp (1-t) \Big] \nnb \\
\ar e_d \Big[\uu (1+t) - \sp (1-t) \Big] -
e_s (\uu+\dd) (1+t) \Big\} \psi^{a\prime}(u_0) \nnb \\
\ek {e\over 288} {1\over M^2} m_0^2
\Big\{ m_d \uu \sp \Big[ 2 e_u ( 5 + 4 t + 5 t^2) +
e_s (1+t)^2\Big] \nnb \\
\ar m_u \dd \sp \Big[ 2 e_d ( 5 + 4 t + 5 t^2) + e_s (1+t)^2\Big] +
2 (e_u + e_d) m_s \uu \dd ( 5 + 4 t + 5 t^2) \Big\}
\widetilde{j}_1(h_\gamma) \nnb \\
\ek {e\over 576} {1\over M^4} \gGgG
\Big\{ m_d \uu \sp \Big[ e_u ( 3 + 2 t + 3 t^2) -
e_s (1+t)^2\Big] \nnb \\
\ar m_u \dd \sp \Big[ e_d ( 3 + 2 t + 3 t^2) - e_s
(1+t)^2\Big] +
(e_u + e_d) m_s \uu \dd ( 3 + 2 t + 3 t^2) \Big\}
\widetilde{j}_1(h_\gamma) \nnb \\
\ek {e\over 1152} {1\over M^6} m_0^2 \gGgG
\Big\{ m_d \uu \sp \Big[ e_u ( 3 + 2 t + 3 t^2) -
e_s (1+t)^2\Big] \nnb \\
\ar m_u \dd \sp \Big[ e_d ( 3 + 2 t + 3 t^2) - e_s
(1+t)^2\Big] +
(e_u + e_d) m_s \uu \dd ( 3 + 2 t + 3 t^2) \Big\}
\widetilde{j}_1(h_\gamma) \nnb \\
\ek {e\over 256 \pi^2} M^4 e_s (1-t)^2 \Big\{
\sp \Big[ 2 i_3({\cal S},1) - 2 i_3(\widetilde{\cal S},1)
- 2 i_3({\cal T}_2,1) + 4 i_3({\cal T}_3,1) - 2 i_3({\cal T}_4,1)
+ 4 i_3({\cal T}_2,v) \nnb \\
\ek 8 i_3({\cal T}_3,v) + 4 i_3({\cal T}_4,v)
- 4 \widetilde{j}_1(h_\gamma) + \mathbb{A} (u_0) \Big] -
f_{3\gamma} (m_u+m_d) \psi^{a\prime} (u_0) \Big\} \nnb \\
\ek {3 e\over 64 \pi^2} M^4 (e_u \uu + m_d \dd) (1-t^2)
\widetilde{j}_1(h_\gamma) \nnb \\
\ar {e\over 256 \pi^2} M^4 f_{3\gamma} (1-t) \Big\{
e_u \Big[ m_d (1+t) - m_s (1-t) \Big] +
e_d \Big[ m_u (1+t) - m_s (1-t) \Big] \nnb \\
\ek 2 e_s (m_u+m_d) \Big\} \psi^{a\prime} (u_0) \nnb \\
\ar {e\over 192 \pi^2} M^6 e_s \sp (1-t)^2 \chi \varphi_\gamma^\prime
(u_0)
\eea

The functions $i_n$, $\widetilde{i}_4$ and
$\widetilde{\widetilde{i}}_4$ are defined as
\bea
\label{nolabel}
i_0(\phi,f(v)) \es \int {\cal D}\alpha_i \int_0^1 dv
\phi(\alpha_{\bar{q}},\alpha_q,\alpha_g) f(v) (k-u_0) \theta(k-u_0)~, \nnb \\
i_1(\phi,f(v)) \es \int {\cal D}\alpha_i \int_0^1 dv
\phi(\alpha_{\bar{q}},\alpha_q,\alpha_g) f(v) \theta(k-u_0)~, \nnb \\
i_2(\phi,f(v)) \es \int {\cal D}\alpha_i \int_0^1 dv
\phi(\alpha_{\bar{q}},\alpha_q,\alpha_g) f(v) \delta(k-u_0)~, \nnb \\
i_3(\phi,f(v)) \es \int {\cal D}\alpha_i \int_0^1 dv
\phi(\alpha_{\bar{q}},\alpha_q,\alpha_g) f(v) \delta^\prime(k-u_0)~, \nnb \\
i_4(\phi,f(v)) \es \int {\cal D}\alpha_i \int_0^1 dv
\phi(\alpha_{\bar{q}},\alpha_q,\alpha_g) f(v) \delta^{\prime\prime}(k-u_0)~, \nnb \\
\widetilde{j}_1(f(u)) \es \int_{u_0}^1 du f(u)~, \nnb \\
\widetilde{j}_2(f(u)) \es \int_{u_0}^1 du (u-u_0) f(u)~, \nnb
\eea
where 
\bea
k = \alpha_q + \alpha_g \bar{v}~,~~~~~u_0={M_1^2 \over M_1^2
+M_2^2}~,~~~~~M^2={M_1^2 M_2^2 \over M_1^2
+M_2^2}~.\nnb
\eea

\end{document}